\documentclass[showpacs,twocolumn,floatfix,superscriptaddress]{revtex4-2}
\usepackage{graphicx}% Include figure files
\usepackage{bm}% bold math
\usepackage[colorlinks=true,urlcolor=blue,citecolor=blue,linkcolor=blue]{hyperref}% add hypertext capabilities
\usepackage{suffix}

\usepackage{xr}
\usepackage{xfrac}
\usepackage{multirow}
\usepackage{amsmath}
\usepackage{amssymb}
\usepackage{braket}
\usepackage{physics}
\usepackage{amsthm}
\usepackage{natbib}
\usepackage{mathtools}
\usepackage[utf8]{inputenc}
\usepackage[english]{babel}
\usepackage{scalerel}[2014/03/10]
\usepackage{stackengine}
\usepackage{quantikz}
\usepackage{algorithm2e}
\usepackage{subfigure}
\RestyleAlgo{ruled}

\SetKwComment{Comment}{/* }{ */}
\makeatother

\newtheorem{theorem}{Theorem}[]
\newtheorem*{remark}{Remark}

\newtheorem{lemma}[]{Lemma}

\newtheorem{definition}{Definition}

\theoremstyle{definition}

\makeatletter
\@addtoreset{proofpart}{theorem}
\@addtoreset{proofcase}{theorem}
\makeatother
\renewcommand{\exp}[1]{\text{exp}\left( #1 \right)}

\newcommand{\supp}[0]{\mathrm{supp}}

\begin{document}
\preprint{APS/123-QED}

\title{Characterizing Pauli Propagation via Operator Complexity}

\author{Yuguo Shao}
\affiliation{Yau Mathematical Sciences Center and Department of Mathematics, Tsinghua University, Beijing 100084, China}
\affiliation{Beijing Institute of Mathematical Sciences and Applications, Beijing 101408, China}
\author{Song Cheng}
\email{chengsong@bimsa.cn}
\affiliation{Beijing Institute of Mathematical Sciences and Applications, Beijing 101408, China}
\author{Zhengwei Liu}
\email{liuzhengwei@mail.tsinghua.edu.cn}
\affiliation{Yau Mathematical Sciences Center and Department of Mathematics, Tsinghua University, Beijing 100084, China}
\affiliation{Beijing Institute of Mathematical Sciences and Applications, Beijing 101408, China}
%\date{\today}

\begin{abstract}
Pauli-propagation simulation represents observables in the Pauli basis and evolves their coefficients in the Heisenberg picture.
Its efficiency depends on whether the evolving operator can be accurately compressed by retaining only a limited number of Pauli terms.
In this work, we bridge operator complexity and the resource cost of Pauli-propagation methods by proving that the truncation error is governed by the Operator Stabilizer Rényi entropy (OSE) $\mathcal{S}^\alpha(O)$.
Our a priori bounds quantify how OSE controls the compressibility of the evolving operator and give explicit prescriptions for the Top-$K$ budget required to achieve a target accuracy.
As an analytic test case, we prove that for the 1D Heisenberg model at $J_z=0$, the number of non-zero Pauli coefficients generated from a local operator grows at most quadratically with the number of Trotter steps.
We then benchmark the Top-$K$ Pauli propagation on XXZ Heisenberg chains.
The numerical results show high accuracy with a small truncation number $K$ in the free regime ($J_z=0$) and competitive performance against tensor-network methods, such as TDVP, in the interacting case ($J_z=0.5$).
These results position OSE as a resource measure for Pauli-propagation methods.

\end{abstract}
\maketitle

% ------------------------ MAIN BODY ------------------------
\section{Introduction}

Simulating real-time quantum dynamics in interacting many-body systems represents a fundamental challenge across condensed matter physics~\cite{cirac2012goals,georgescu2014quantum,polkovnikov2011colloquium,eisert2015quantum}, quantum information science~\cite{miessen2023quantum,daley2022practical,fauseweh2024quantum,childs2022quantum,liao2023simulation}, and statistical mechanics~\cite{weimer2021simulation,d2016quantum,gogolin2016equilibration,vasseur2016nonequilibrium,mori2018thermalization}. 
Understanding non-equilibrium phenomena such as thermalization~\cite{abanin2019colloquium,mori2018thermalization,d2016quantum,gogolin2016equilibration}, information scrambling~\cite{swingle2018unscrambling,mi2021information,zanardi2021information}, and transport properties requires accurate numerical methods for evolving quantum states or observables in time~\cite{waintal2024computational,shevtsov2013numerical}. 

However, the exponential growth of Hilbert space dimension severely constrains classical simulations~\cite{feynman2018simulating,feng2025quon,kang20252d,xu2023herculean}.
Traditional approaches face critical limitations: Exact diagonalization is restricted to small systems by memory constraints, while tensor network methods encounter barriers with the rapid growth of entanglement entropy in generic quenched dynamics~\cite{biamonte2017tensor,orus2019tensor,orus2014practical,verstraete2006matrix,paeckel2019time}.
This entanglement barrier fundamentally restricts the simulation timescales for density matrix renormalization group (DMRG)~\cite{schollwock2005density,schollwock2011density,qian2024augmenting} and time-dependent variational principle (TDVP)~\cite{haegeman2016unifying,haegeman2011time,qian2025clifford} approaches, particularly in higher dimensions or in systems undergoing information scrambling~\cite{vznidarivc2020entanglement,nahum2017quantum,erhard2020advances,paeckel2019time,wild2023classical,wu2025classicalalgorithmshamiltoniandynamics}.

To circumvent the entanglement bottleneck, recent advances have revisited the Heisenberg picture, focusing on the evolution of operators rather than states.
This operator-centric perspective has gained traction through Pauli-propagation-based approaches for simulating noisy quantum circuits~\cite{aharonov2023polynomial,schuster2024polynomial,gao2018efficient,shao2024simulating,angrisani2024classically,angrisani2025simulating,fontana2025classical,rudolph2023classical,shao2025diagnosing,martinez2025efficient,zhang2024clifford,beguvsic2025simulating,beguvsic2025real,lerch2024efficient,rudolph2025pauli,lin2025utility}, such as the Observable's Back-Propagation Pauli Propagation (OBPPP)~\cite{shao2024simulating} or Low Weight Efficient Simulation Algorithm (LOWESA)~\cite{fontana2025classical,rudolph2023classical}, et al.
However, while these methods have demonstrated utility in circuit simulations, a rigorous theoretical framework connecting the computational complexity of Hamiltonian dynamics to intrinsic operator properties remains under-explored.
In practice, these methods become useful only when the evolving operator remains compressible in the Pauli basis, so that one can truncate the expansion without losing the desired accuracy.

This raises a resource-theoretic question for Pauli propagation: what intrinsic property of an operator determines the number of Pauli terms that must be retained?
Truncation rules such as largest-coefficient~\cite{loizeau2025quantum,Loizeau2025}, threshold~\cite{beguvsic2025simulating,beguvsic2025real}, or weight truncation~\cite{aharonov2023polynomial,shao2024simulating,fontana2025classical,schuster2024polynomial} are algorithmic choices, but their accuracy depends on the distribution of Pauli coefficients generated during the evolution.
Thus, beyond proposing a particular truncation scheme, one needs a complexity measure that connects this coefficient distribution to quantitative error guarantees.

In this work, we bridge operator complexity and the simulation of quantum spin systems by establishing a rigorous link between Pauli propagation accuracy and the \textit{Operator Stabilizer Rényi entropy (OSE)}~\cite{dowling2025magic}.
Conceptually, the OSE serves as the operator-space dual to the Rényi entropies used for states~\cite{leone2022stabilizer}.
Our primary contribution is the derivation of \textit{a priori} error bounds for amplitude truncation, proving that the simulation error is explicitly controlled by the OSE of the evolving operator.
Equivalently, for a target error tolerance, these bounds prescribe how the required Pauli-path budget scales with OSE.
Just as entanglement entropy quantifies the difficulty of tensor-network simulations, we demonstrate that OSE quantifies the difficulty of Pauli propagation methods.
We provide explicit resource prescriptions, showing that the number of Pauli terms required to maintain a target accuracy scales with the exponent of the OSE.
This formalizes the intuition that observables with low ``magic'' or ``non-stabilizerness'' remain highly compressible, allowing for efficient classical simulation even when state entanglement is high.

Complementing this general bound, we prove a structure theorem for the paradigmatic 1D XY Heisenberg chain ($J_z = 0$).
We analytically show that the number of non-zero Pauli coefficients generated from a local operator $Z_l$ scales quadratically with the number of Trotter steps, i.e., $\order{s^2}$.
This explains the sustained effectiveness of aggressive truncation even at late times and provides a concrete example where operator-complexity-based approaches could outperform those based on state entanglement.
It highlights a regime where operator-complexity-based approaches fundamentally outperform those limited by state entanglement.

To validate these theoretical insights numerically, we employ an explicit algorithmic realization: the Observable's Back-Propagation Pauli Propagation with a Top-$K$ truncation strategy.
For convenience, we shorten this approach as OBPPP with Top-$K$ truncation.

This method retains the $K$ dominant Pauli coefficients at each time step, followed by norm rescaling for stability, directly leveraging the compressibility predicted by our theory. We prove that the error incurred by the Top-$K$ truncation strategy decays at a rate controlled by OSE. This provides explicit resource prescriptions for choosing a proper $K$ to meet a target accuracy. Conceptually, the OSE plays the role dual to the Stabilizer Rényi entropies used for states~\cite{leone2022stabilizer}, formalizing the intuition that observables with low OSE remain highly compressible under Heisenberg evolution.
An identical implementation has also been independently developed in PauliStrings.jl package~\cite{loizeau2025quantum,Loizeau2025}.

We benchmark this framework on 1D Heisenberg chains.
In free regimes ($J_z = 0$), the method achieves high accuracy with a small $K$, consistent with our structure theorem.
In interacting regimes ($J_z = 0.5$), where OSE grows more rapidly, the method remains competitive with state-of-the-art tensor-network methods (e.g., TDVP), offering a viable alternative when entanglement barriers are prohibitive.
These numerical results support the central interpretation that OSE quantifies the computational resource required by Pauli-propagation simulations.

Another related exploration on Pauli propagation in Hamiltonian dynamics has been reported in Ref.~\cite{beguvsic2024fast,beguvsic2025simulating,beguvsic2025real}, where the authors study two variants of Pauli-propagation-based methods, namely Sparse Pauli Dynamics (SPD) and X-truncated SPD (xSPD), and benchmark them on 1D spin chains, 2D and 3D transversal-field Ising model. The main difference between our work and Ref.~\cite{beguvsic2025real} lies in the truncation strategy, where we consider Top-$K$ truncation with norm rescaling while they use a fixed threshold truncation without rescaling. More comparison of those works are summarized in Appendix~\ref{ap:comparison}.

The paper is organized as follows: Sec.~\ref{sec:method} details OBPPP with Top-$K$ truncation and algorithmic implementation, Sec.~\ref{sec:error_bounds} presents analytical error bounds based on OSE, Sec.~\ref{sec:numerical_benchmarks} contains numerical benchmarks on Heisenberg models, and Sec.~\ref{sec:conclusions} provides conclusions and future directions.

\section{OBPPP with Top-$K$ Truncation}\label{sec:method}
Consider $d\rho/dt = -i[H,\rho]$ with a Pauli decomposition
$H=\sum_{i=1}^{N_H} w_i P_i,\; P_i \in \{\mathbb{I},X,Y,Z\}^{\otimes n}$, where $N_H$ is the number of Pauli words $P_i$ and $n$ is the system size, $w_i$ are the Pauli coefficients.
The density matrix evolves as
\begin{equation}
  \rho(t) = \exp{-iHt}\,\rho(0)\,\exp{iHt}.
\end{equation}
Using the Trotter decomposition with $\tau=t/N$, there is:
\begin{equation}
  \exp{-iHt} = \prod_{s=1}^{N} \prod_{i=1}^{N_H} \exp{-i w_i P_i \tau} + \mathcal{O}(\tau^2).
\end{equation}
Thus, the density matrix at time $t$ can be approximated as:
\begin{equation}
  \rho(t) \approx \left[\prod_{s=1}^{N} U(\tau) \right] \rho(0) \left[\prod_{s=1}^{N} U(\tau) \right]^{\dagger},
\end{equation}
where $U(\tau):=\prod_{i=1}^{N_H} \exp{-i w_i P_i \tau}$ is the Trotterized evolution operator for a single time step.

For an observable $O = \sum_j c_j P_j$, we could calculate the expectation value $\langle O \rangle_{\rho(t)} = \tr{O \rho(t)}$ by back-propagating those Pauli words $P_j$ in the Heisenberg picture under Pauli evolution.

Let $\vec{O}=(O_0, O_1, \ldots, O_N)$ denote the evoluted trajectory of $O$ under reverse order of time. Therefore, $O_N=O$ and $O_{N-1}, O_{N-2}, \ldots, O_0$ defined recursively. That is, for step $s$, we have $O_s = \sum_{j} c_{s,j} P_{s,j}$ with $c_{s,j}=\frac{1}{2^n}\tr{P_{s,j} O_s}$, then the next operator:
\begin{equation}\label{eq:observable_back_propagation}
  \begin{aligned}
    O_{s-1} &= U(\tau)^\dagger O_{s} U(\tau) \\
    &= \sum_{j} c_{s,j} U(\tau)^\dagger P_{s,j} U(\tau).
  \end{aligned}
\end{equation}
Consequently, calculating the expectation value of $O$ at time $t$ is now related to calculating the final operator $O_0$:
\begin{equation}
  \langle O \rangle_{\rho(t)} \approx \tr{O \prod_{s=1}^{N} U(\tau) \rho(0) \prod_{s=1}^{N} {U(\tau)}^\dagger} = \tr{O_0 \rho(0)},
\end{equation}

For a single pair of Pauli words $(P_{s,j}, P_i)$, there is a closed-form~\cite{zhang2024clifford,beguvsic2025real,shao2024simulating}:
\begin{small}
  \begin{equation}\label{eq:pauli_evolution}
    \begin{aligned}
    &\left[\exp{-i w_i P_i \tau}\right]^\dagger P_{s,j} \left[\exp{-i w_i P_i \tau}\right]\\
    &= 
    \begin{cases}
      P_{s,j}, & \text{if } [P_i, P_{s,j}] = 0  \\
      \cos(2w_i \tau) P_{s,j} - i\sin(2w_i \tau) P_i P_{s,j}, & \text{if } \{P_i, P_{s,j}\} = 0 
      \end{cases},
    \end{aligned}
  \end{equation}
\end{small}
which can be calculated in $\order{n}$ by checking the commutation relation.

However, naively employing Eq.~\eqref{eq:pauli_evolution} to Eq.~\eqref{eq:observable_back_propagation} would cause $\abs{O_{s-1}}$ to grow combinatorially as $\order{2^{N_H}\abs{O_s}}$, where $\abs{O_s}$ denotes the number of terms in $O_s$, making the computation infeasible.

To address this issue, we consider the Top-$K$ truncation strategy at each step:
given a threshold $K$, retain only $K$ coefficients with largest magnitude in $O_s=\sum_{j} c_{s,j} P_{s,j}$ and define:
\begin{equation}
  \widehat{O}_s = \sum_{j=1}^{K} c_{s,j}' P_{s,j}',
\end{equation}
where $\{c_{s,j}'\}_{j=1}^{K}$ are the selected $K$ coefficients and $\{P_{s,j}'\}_{j=1}^{K}$ are the associated Pauli words.
We denote the resulting truncated observable sequence by $\{\widehat{O}_s\}_{s=0}^{N}$.
Under this scheme, the total cost is $\order{n N N_H K}$.

Additionally, after truncation, we rescale $\widehat{O}_0$ to match the Hilbert-Schmidt norm of $O_0$:
\begin{equation}
  \widehat{O}_0 \gets \widehat{O}_0 \times \frac{\norm{O}_2}{\norm{\widehat{O}_0}_2},
\end{equation}
such that $\norm{\widehat{O}_0}_2 = \norm{O_0}_2$.
Finally, we approximate the $\langle O \rangle_{\rho(t)}$ by $\langle \widehat{O} \rangle_{\rho(t)} = \tr{\widehat{O}_0 \rho(0)}$.

The above procedure is summarized as Algorithm~\ref{alg:algorithm} and this pipeline has also been implemented in PauliStrings.jl package~\cite{loizeau2025quantum,Loizeau2025}.
In our implementation, exact sorting is not required: an approximate Top-$K$ truncation can be realized by a bucket-based selection strategy with $\order{1}$ bucket assignment cost, so the sorting step does not introduce additional asymptotic overhead.
More details are provided in Appendix~\ref{ap:computational_cost}.

In addition to Top-$K$ truncation, another common truncation strategy in Pauli-based simulations is Hamming weight truncation~\cite{aharonov2023polynomial,schuster2024polynomial,shao2024simulating,angrisani2024classically,angrisani2025simulating,fontana2025classical,martinez2025efficient,gonzalez2025pauli}, which retains only Pauli terms with lower Hamming weight. We also investigate this strategy and find that it is unnecessary in Heisenberg model simulations, see Appendix~\ref{ap:weight_truncation}.

\begin{algorithm}[h]\label{alg:algorithm}
  \caption{OBPPP with Top-$K$ Truncation for Real-time Evolution}
  \KwIn{Hamiltonian $H$, initial state $\rho(0)$, observable $O$, time $t$, number of time steps $N$, term budget $K$}
  \KwOut{Approximated observable value $\langle \hat{O} \rangle_{\rho(t)}$}

  \textbf{Set} $\tau \gets t/N$\;
  \textbf{Set} $\widehat{O}_N \gets O$\;
  
  \For{$s$ from $N$ to $1$}{
    \textbf{Set} $\widehat{O}_{s-1} \gets \widehat{O}_s$\;
    
    \For{$i$ in $1, \ldots, N_H$}{
      \textbf{Set} $O_{\text{Temp}} \gets 0$\;
      
      \ForAll{Pauli terms $P_\alpha$ in $\widehat{O}_{s-1}=\sum_\alpha c_\alpha P_\alpha$}{
        \textbf{Set} $O_{\text{Temp}} \gets O_{\text{Temp}} + \exp{-i w_i P_i \tau}^\dagger P_\alpha \exp{-i w_i P_i \tau}$\;
      }

      \textbf{Set} $O_{\text{Temp}} \gets \text{Top-}K \text{ terms by } |c_\alpha| \text{ from } O_{\text{Temp}}$\;

      \textbf{Set} $\widehat{O}_{s-1} \gets O_{\text{Temp}}$\;
    }
    
  }
  
  \textbf{Set} $\widehat{O}_0 \gets \widehat{O}_0 \times \frac{\norm{O}_2}{\norm{\widehat{O}_0}_2}$\;
  \textbf{Set} $\langle \hat{O} \rangle_{\rho(t)} \gets \tr{\widehat{O}_0 \rho(0)}$\;
  \Return{$\langle \hat{O} \rangle_{\rho(t)}$}\;
\end{algorithm}

\section{Analytical Bounds}\label{sec:error_bounds}
In this section, we provide analytical bounds on the error of the Top-$K$ truncation strategy.
Firstly, we utilize the OSE, an intrinsic measure of the complexity of an operator in the Pauli basis.

\begin{definition}[Operator Stabilizer Rényi entropy~\cite{dowling2025magic}]\label{def:pauli-operator-renyi-entropy}
  Let $O$ be an operator on $n$ qubits with Pauli coefficients $c_i=\frac{1}{2^n}\tr{P_i O}$ for $P_i \in \{I,X,Y,Z\}^{\otimes n}$, and denote the squared coefficients by $\bm{c^2}=(c_1^2,c_2^2,\ldots)$. For $\alpha>0$, $\alpha\neq 1$, the Operator Stabilizer Rényi entropy~(OSE) of $O$ is defined as:
  \begin{equation}
    \mathcal{S}^\alpha(O) \coloneqq \frac{\alpha}{1-\alpha}\,\ln{\norm{\bm{c^2}}_\alpha},
  \end{equation}
  where $\norm{\bm{c^2}}_\alpha=\left(\sum_{i=1}^\infty c_i^{2\alpha}\right)^{\frac{1}{\alpha}}$.
\end{definition}

This quantity can be viewed as the adjoint analogue of the stabilizer Rényi entropy~\cite{leone2022stabilizer}, acting on observables rather than states.
Like its state counterpart, OSE quantifies the ``non-stabilizerness'' or ``magic'' of an operator.
A resource theory framework is established for OSE~\cite{dowling2025magic}.
The following lemma records several basic properties of OSE, where properties (1)-(2) were established in Ref.~\cite{dowling2025magic}:
\begin{lemma}\label{lemma:prop_entropy}
  The OSE $\mathcal{S}^\alpha(O)$ satisfies the following properties:
  \begin{enumerate}
    \item $\mathcal{S}^\alpha(O_1 \otimes O_2) = \mathcal{S}^\alpha(O_1) + \mathcal{S}^\alpha(O_2)$.
    \item For a Clifford circuit $U$, we have $\mathcal{S}^\alpha(U O U^\dagger) = \mathcal{S}^\alpha(O)$.
    \item For $0 < \alpha < 0.5$, $\mathcal{S}^\alpha(O)$ is concave, i.e., for any $\sum_i \lambda_i = 1$ with $\lambda_i \geq 0$, we have $\mathcal{S}^\alpha(\sum_i \lambda_i O_i) \geq \sum_i \lambda_i \mathcal{S}^\alpha(O_i)$.
    \item For $0.5 \leq \alpha <1 $, $\exp{\mathcal{S}^\alpha(O)}$ is convex, i.e., for any $\sum_i \lambda_i = 1$ with $\lambda_i \geq 0$, we have $\exp{\mathcal{S}^\alpha(\sum_i \lambda_i O_i)} \leq \sum_i \lambda_i \exp{\mathcal{S}^\alpha(O_i)}$.
  \end{enumerate}
\end{lemma}

We now establish the following error bound for the Top-$K$ truncation, where the lower bound part is initially derived in Ref.~\cite{dowling2025magic}:
\begin{lemma}\label{lemma:error_bound}
  Let $O = \sum_{P \in \{I, X, Y, Z\}^{\otimes n}} c_P P$, where $c_P = \frac{1}{2^n} \tr{P O}$. The approximated observable is given by:
  \begin{equation}
    \widehat{O} = \frac{\norm{O}_2}{\sqrt{2^n\sum_{i=1}^{K} c'^2_i}}\sum_{i=1}^{K} c'_i P'_i,
  \end{equation}
  where $\{c'_i\}$ and $\{P'_i\}$ denote the $K$ largest magnitude coefficients and their corresponding Pauli words, respectively.
  Then the truncation error is bounded by:
\begin{equation}\label{eq:error_bound}
  \sqrt{\varDelta(K)} \leq \norm{O-\widehat{O}}_{Pauli,2} \leq \sqrt{2\varDelta(K)}.
\end{equation}
where $\varDelta(K) \coloneqq \sum_{i=K+1}^{\infty} c'^2_i$ denotes the squared sum of the tail terms. $\norm{A}_{Pauli,2} = \sqrt{\frac{1}{2^n}\tr{A^\dagger A}}$ is the normalized Hilbert-Schmidt norm.
\end{lemma}

In fact, the lower bound obtained in Ref.~\cite{dowling2025magic} relates to the $\infty$-norm rather than the $2$-norm, the relation between those and limitations are discussed in Appendix~\ref{ap:proof_error_bound}.

For matrix product states~(MPS), Ref.~\cite{verstraete2006matrix} showed the connection between the truncation error for given bond dimension and Rényi entropy, which has been the foundation of the error analysis for tensor-network methods.
Here, building on the preceding lemma, we can similarly bound $\varDelta(K)$ using the OSE $\mathcal{S}^\alpha(O)$:
\begin{theorem}\label{thm:sim_error_entropy}
  Let $O$ be an observable, for $0 < \alpha < 1$, we have:
  \begin{equation}
    \ln{\varDelta(K)} \leq \frac{1-\alpha}{\alpha}\left( \mathcal{S}^\alpha(O)-\ln{K} \right) + \ln{\frac{\alpha}{1-\alpha}}.
  \end{equation}
\end{theorem}

Consequently, for a given target error $\varepsilon$, to ensure $\norm{O-\widehat{O}}_{Pauli,2} \leq \varepsilon$, it suffices to require $\varDelta(K) \leq \frac{\varepsilon^2}{2}$. 
By Theorem~\ref{thm:sim_error_entropy}, this condition is satisfied whenever:
\begin{equation}
  K \geq \exp{\mathcal{S}^\alpha(O) } \left(\frac{2\alpha}{(1-\alpha)\varepsilon^2}\right)^{\frac{\alpha}{1-\alpha}}
\end{equation}

As a complementary result in the opposite direction, Ref.~\cite{dowling2025magic} established a lower bound on $\varDelta(K)$ for the case of $\alpha=1$, here we consider the case for general $\alpha$:
\begin{theorem}\label{thm:lower_bound}
  Let $O$ be an observable, for $0 < \alpha < 1$, we have:
  \begin{small}
    \begin{equation}\label{eq:lower_bound}
      \begin{split}
      \mathcal{S}^\alpha(O) &\leq \frac{1}{1-\alpha} \ln\Big( \left( \norm{O}_{\mathrm{Pauli},2}^2 - \varDelta(K) \right)^{\alpha} K^{1-\alpha} \\
      &\quad + \varDelta(K)^{\alpha} (4^n - K)^{1-\alpha} \Big),
    \end{split}
  \end{equation}
  \end{small}
  while for $\alpha > 1$, we have:
  \begin{equation}
    \mathcal{S}^\alpha(O) \leq \ln{K} - \frac{\alpha}{\alpha-1} \ln{\left( \norm{O}_{\mathrm{Pauli},2}^2 - \varDelta(K) \right)},
  \end{equation}
  where the right-hand sides in both cases are monotonic increasing with respect to $\varDelta(K)$ in its domain, and thus implicitly provides a lower bound on $\varDelta(K)$ for given $\mathcal{S}^\alpha(O)$ and $K$.
  
  In particular, when $\alpha \to 1_-$, Eq.~\eqref{eq:lower_bound} gives the result in Ref.~\cite{dowling2025magic}:
  \begin{equation}
    \varDelta(K) \geq \frac{1}{2n} \left(\frac{1}{\ln{2}}\mathcal{S}^1(O) - \log_2 K -1 \right).
  \end{equation}
\end{theorem}

Here, we mainly focus on the case of the upper bound of complexity, such as Thm.~\ref{thm:sim_error_entropy}, which provides practical guidance on how to choose $K$ to achieve a target accuracy.
For the lower bound, Thm.~\ref{thm:lower_bound} provides complementary insights into the fundamental limits of compressibility, and there are some specific examples discussed in Ref.~\cite{dowling2025magic}.

\section{1D Heisenberg system}\label{sec:numerical_benchmarks}

The one-dimensional Heisenberg model is a paradigmatic quantum many-body system comprising spin-1/2 degrees of freedom on a one-dimensional chain and coupled via nearest-neighbor exchange interactions:
\begin{equation}
  H
  =  \sum_{i=1}^{L-1}
  J_x \sigma_i^x \sigma_{i+1}^x + J_y \sigma_i^y \sigma_{i+1}^y + J_z \sigma_i^z \sigma_{i+1}^z ,
\end{equation}
where $L$ is the chain length, $J_{x,y,z}$ are the coupling constants, and $\sigma_i^{x,y,z}$ are Pauli operators acting on site $i$. 

Like low-entangled structures for tensor network representations, the key to controlling the complexity of the Pauli propagation approach is to find systems with low OSE.
There are two specific cases where the OSE can be analytically bounded. The first is the Clifford (or ``Low magic'') circuits, where the OSE remains constant~\cite{bravyi2019simulation,lerch2024efficient}.
The second is the case that the dimension of the Dynamical Lie Algebra (DLA) generated by $H$ and the observable $O$ is of polynomial size~\cite{goh2023lie}.
For example, Ref.~\cite{goh2023lie} shows that the DLA dimension for the 1D Transverse Field Ising Model (TFIM) scales as $\order{n^2}$, thus the OSE grows at most logarithmically with system size, leading to efficient Pauli propagation.
Here, we consider another important example, the Heisenberg model in the case of $J_z=0$, which can be mapped to a free-fermion system~\cite{lieb1961two,katsura1962statistical,prosen2007operator}.
In this case, we prove a structure theorem showing that the time evolution of physical operator $Z_l$ admits a compact representation in the Pauli basis. 
In particular, we prove that:
\begin{theorem}\label{thm:1d_XY_trotter}
  For Hamiltonian given by:
  \begin{equation}
    H = \sum_{i=1}^{n} \left( J_X \sigma_i^x \sigma_{i+1}^x + J_Y \sigma_i^y \sigma_{i+1}^y \right),
  \end{equation}
  the corresponding $s$-step Trotterized evolution operator is:
  \begin{equation}
    U_s = \prod_{t=1}^{s} \prod_{i=1}^{n} \exp{-i \left( J_X \sigma_i^x \sigma_{i+1}^x + J_Y \sigma_i^y \sigma_{i+1}^y \right) \tau}.
  \end{equation}
  Then the evolved operator $Z_l$ satisfies:
  \begin{equation}
    \abs{\left\{ c_p \mid c_p = \tr{P U_s^\dagger Z_l U_s} \neq 0 \right\} } = \order{s^2},
  \end{equation}
  for any $l \in \{1, \ldots, n\}$ and $ P \in \{I, X, Y, Z\}^{\otimes n}$.
  In other words, after $s$-step Trotterization, the number of non-zero Pauli coefficients in the evolved operator $U_s^\dagger Z_l U_s$ scales as $\order{s^2}$.
  Therefore, the OSE of $U_s^\dagger Z_l U_s$ is upper bounded by $\mathcal{S}^\alpha(U_s^\dagger Z_l U_s) \leq \frac{\alpha}{1-\alpha} \ln{\order{s^2}} = \order{\ln{s}}$.
\end{theorem}

This compact structure at $J_z=0$, predicted by the theory, manifests as at-most-quadratic computational complexity for Pauli propagation. 
For $J_Z \neq 0$, related work on the Heisenberg-evolved operators has been reported in Ref.~\cite{dowling2025magic}, where the authors study the OSE of 1D dual unitary XXZ circuits~(can be compiled into SWAP gates and $R_{ZZ}$ gates) for local unitary observables. They found that the OSE can grow linearly with time.

To demonstrate this, we examine two parameter regimes of the Heisenberg model, $J_x=J_y=1, J_z=0$ and $J_x=J_y=1, J_z=0.5$, and compare the OBPPP with Top-$K$ truncation with the tensor-network TDVP method implemented in the ITensor.jl package~\cite{ITensor}.

We consider a chain of length $L=50$ under open boundary conditions~(OBC).
The system is initialized in the Neel state $\ket{\psi} = \ket{\uparrow \downarrow \uparrow \downarrow \ldots}$, and we measure the staggered magnetization $m_z(t) = \frac{1}{L} \sum_{i=1}^{L} (-1)^i \langle S^z_i(t) \rangle$. 
The evolution is simulated up to time $t=10$ with a time step $\tau=0.05$.

\begin{figure}[htbp]
  \centering
  \includegraphics[width=0.5\textwidth]{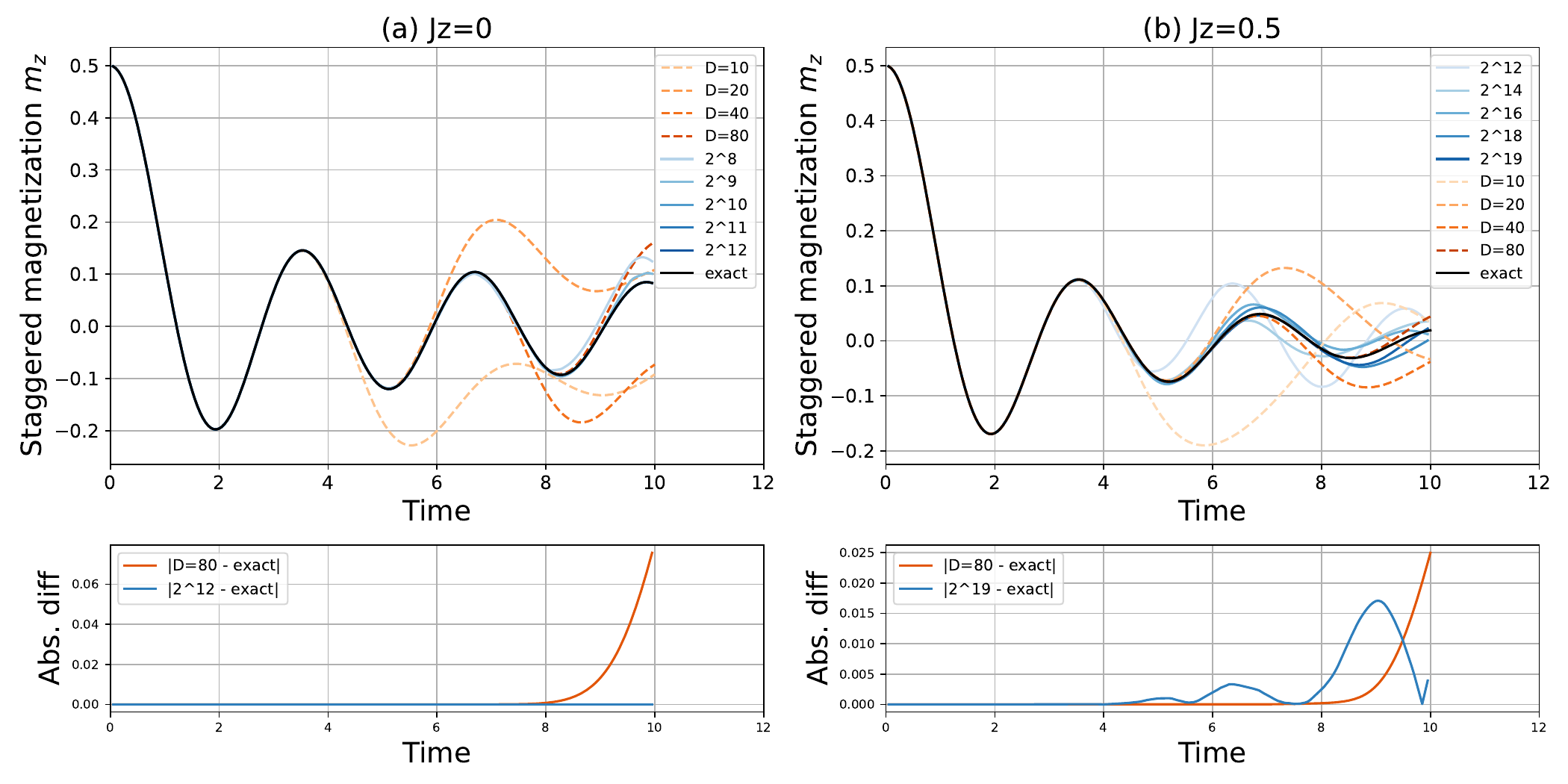}
  \caption{
  Time-evolution results for the one-dimensional Heisenberg model with length $L=50$, obtained using the Pauli propagation and TDVP methods, respectively. The orange lines denote TDVP results for different maximum bond dimensions $D$, and the blue lines denote Pauli propagation results for different $K$ values. 
  For one representative curve, we also compare the absolute deviations from the exact results. 
  (a) For $J_x=J_y=1$ and $J_z=0$, Pauli propagation attains accurate results at tiny cost, whereas for TDVP the error accumulates rapidly once the entanglement entropy exceeds the MPS capacity. (b) For $J_z \neq 0$, TDVP behaves similarly, while the computational cost of Pauli propagation is strongly affected by $J_z$ and becomes comparable to TDVP.
  }
  \label{fig:pauli_vs_tdvp}
\end{figure}

As shown in Fig.~\ref{fig:pauli_vs_tdvp}(a), when $J_z=0$, the Pauli propagation method attains accurate results with a very small Top-$K$ budget of $2^{12}$. In contrast, for TDVP, once $t>8$ the entanglement entropy exceeds the MPS capacity and the error accumulates rapidly.
As shown in Fig.~\ref{fig:pauli_vs_tdvp}(b), when $J_z\neq 0$, TDVP behaves similarly, whereas the computational cost of Pauli propagation is strongly affected by $J_z$ and the accuracy becomes comparable to TDVP. Note that an MPS with bond dimension $D=80$ and Pauli propagation with $K=2^{19}$ have comparable numbers of free real parameters (about 730,000 for MPS and 512,000 for Pauli propagation).

Additionally, the exact reference results in the figure are obtained using TDVP without any bond dimension truncation.
Therefore, we disregard Trotter errors and focus on comparing different representations as well as the impact of the corresponding truncation strategies.

\begin{figure}[htbp]
  \centering
  \includegraphics[width=0.5\textwidth]{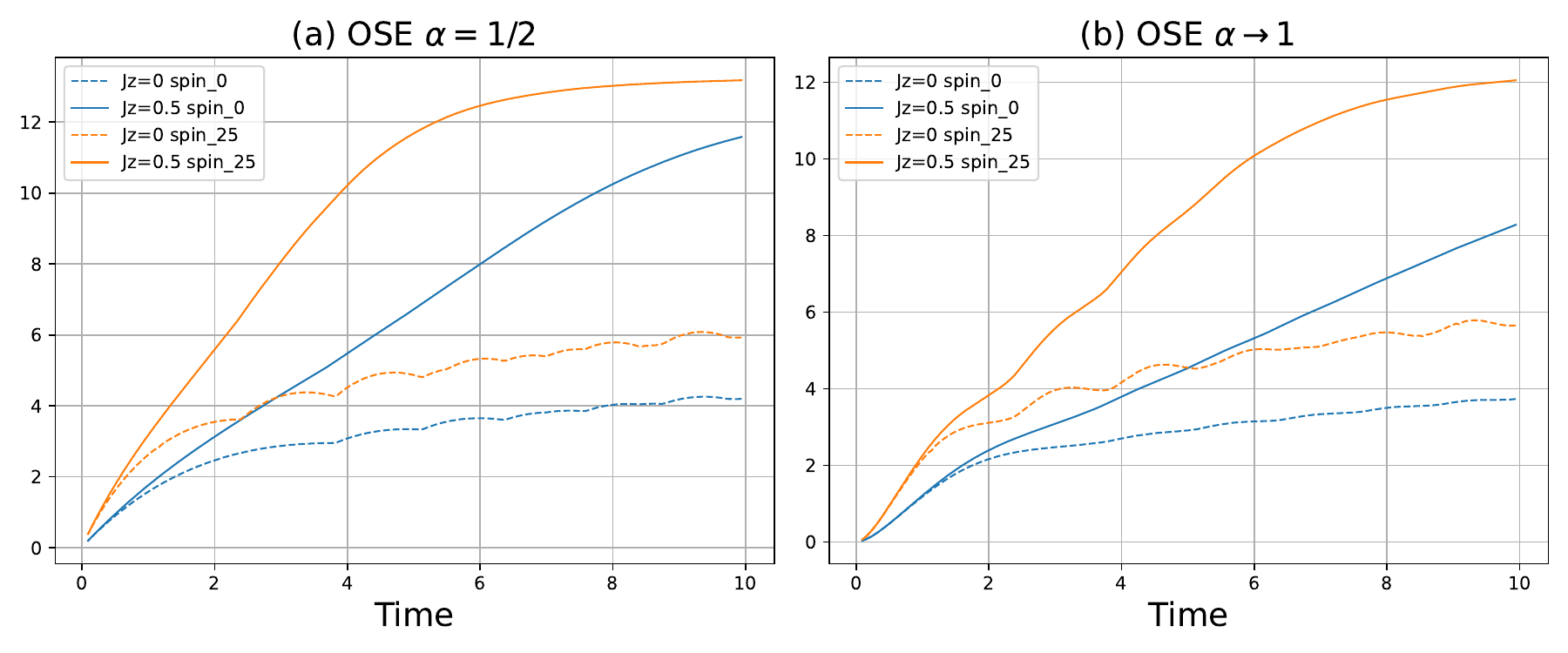}
  \caption{$\alpha = 1/2$ and $\alpha \rightarrow 1$(Shannon) OSE over time. Where the blue line denotes the entropy of the Pauli Z operator with time evolution on the open boundary, and the orange line denotes the entropy corresponding to the Pauli Z operator with time evolution at the center of the one-dimensional chain.
  }
  \label{fig:entropy}
\end{figure}

In Fig.~\ref{fig:entropy}, we plot the OSE at $\alpha = 1/2$ and $\alpha \rightarrow 1$~(Shannon) over time.
In both cases, the entropy increases with time, indicating that the operator becomes more and more complex with time evolution.
Moreover, for a fixed time the entropy increases with $J_z$, reflecting greater operator complexity as $J_z$ increases.

As shown in Theorem~\ref{thm:sim_error_entropy}, once the target tolerance $\varepsilon$ is given, the required value of $K$ is entirely determined by the OSE. 
As an example, consider the propagation of the Pauli Z operator at the chain center: the maximum value of $\mathcal{S}^{\frac{1}{2}}(O)$ is $6.08$ for $J_z=0$ and $13.18$ for $J_z=0.5$, respectively.
While setting $\varepsilon = 0.001$, a sufficient condition to meet this error is $K > 8.7 \times 10^8$ for $J_z= 0$ and $K>1.1 \times 10^{12} $ for $J_z=0.5$. 
Note that this bound is a loose theoretical estimate. 
In practice, our numerical results in Fig.~\ref{fig:pauli_vs_tdvp}(a) show that it achieves accurate results once $K>4\times10^4$. 
Additionally, we indeed numerically observed a higher cost when entropy is larger in the case of $J_z \neq 0$.

\begin{figure}[htbp]
  \centering
  \includegraphics[width=0.5\textwidth]{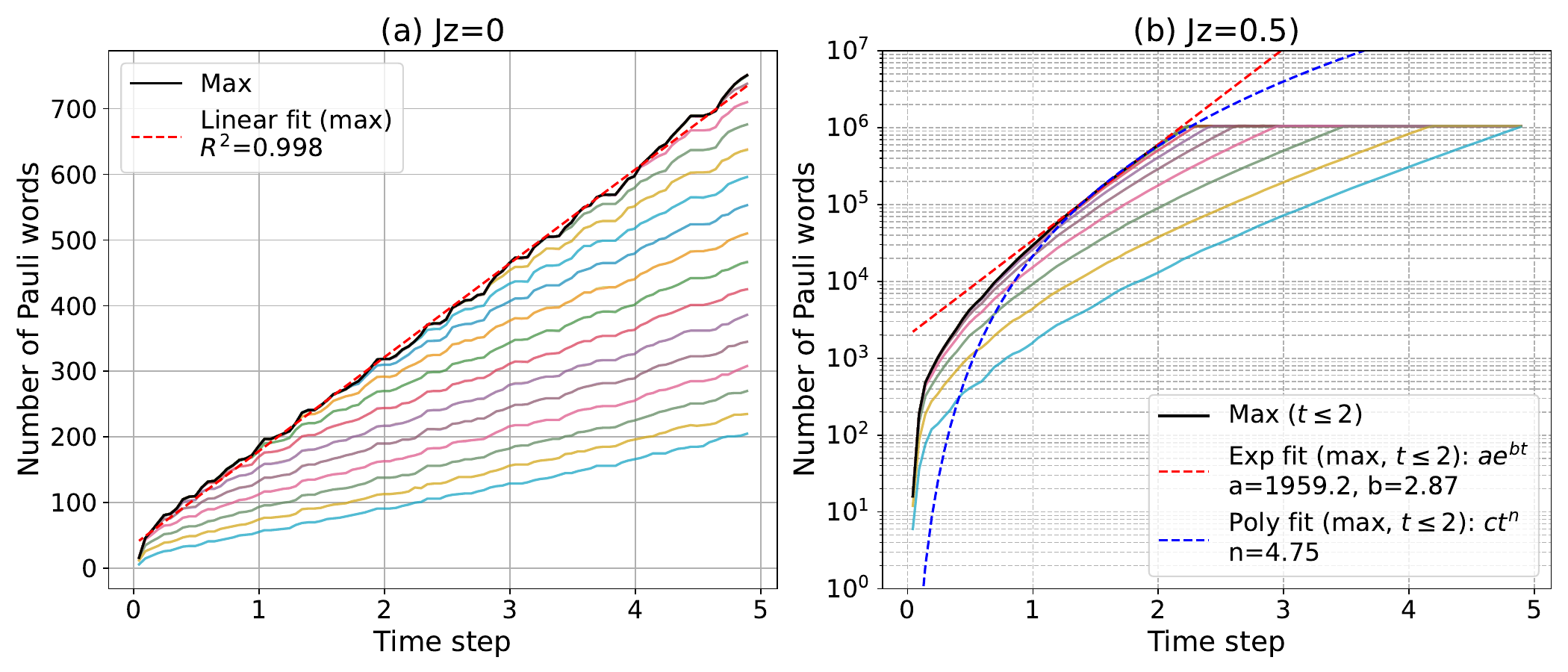}
  \caption{Growth behavior of Pauli words over time for different $J_z$ values. 
  }
  \label{fig:pauli_growth}
\end{figure}

\begin{figure}[htbp]
  \centering
  \includegraphics[width=0.5\textwidth]{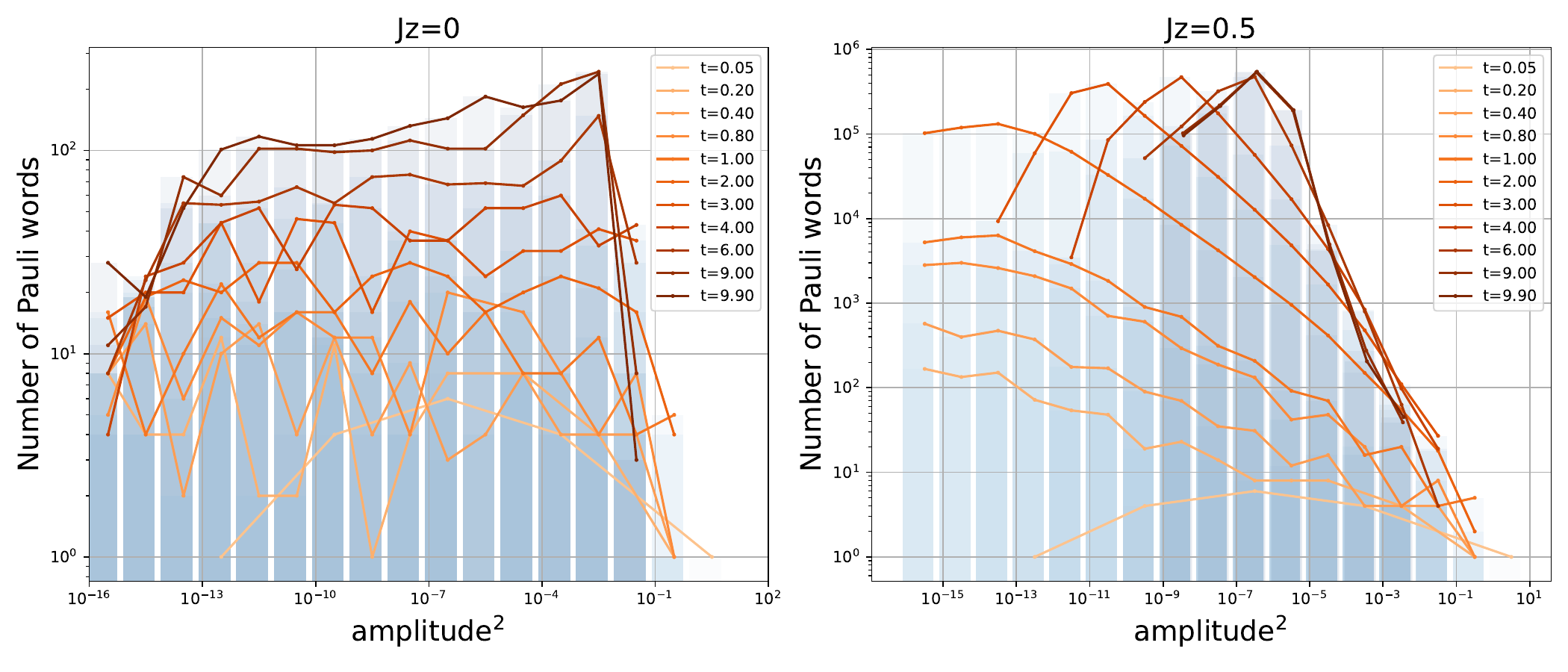}
  \caption{Distribution of the squared Pauli coefficients as a function of time for different values of $J_z$.
  }
  \label{fig:dist}
\end{figure}

In Fig.~\ref{fig:pauli_growth}, we plot the growth behavior of Pauli operators with non-zero coefficients over time for different $J_z$ values.
For $J_z = 0$, the number of Pauli coefficients grows almost exclusively linearly in time. This linear scaling implies that efficiently exact simulation of the dynamic evolution using Pauli propagation method is possible, which is consistent with our theoretical analysis.

While for $J_z = 0.5$, the growth accelerates significantly and shows a quadratic growth trend.
In Fig.~\ref{fig:dist}, we plot the distribution of the squared Pauli coefficients as a function of time for different values of $J_z$.
We can see that the distribution of the squared Pauli coefficients becomes more dispersed with time, and the dispersion becomes more significant with the increase of $J_z$.

\section{Conclusions and Discussions}\label{sec:conclusions}

In this work, we analyzed a Pauli propagation framework for simulating the real-time evolution of local observables in quantum spin systems. The core of this method is to efficiently back-propagate the target observable through elementary Pauli evolutions with a computationally feasible Top-$K$ truncation strategy that retains only the most significant Pauli coefficients at each evolution step.

For analytical error bound, we have derived rigorous a priori bounds that explicitly link the truncation accuracy to OSE. We derived the relationship between truncation threshold $K$ and the OSE, establishing that higher OSE values necessitate larger $K$ for accurate simulations. Theorem~\ref{thm:sim_error_entropy} provides a prescription for choosing the truncation threshold K to achieve a desired accuracy $\varepsilon$, establishing that the simulation cost is governed by the intrinsic operator complexity rather than the entanglement of the quantum state. Pauli propagation approach circumvents the entanglement barriers that challenge state-based tensor network methods.

We further demonstrated the compressibility of Heisenberg-evolved operators through an analytical study of the 1D XY model ($J_z = 0$). We showed that the number of non-zero Pauli coefficients generated from a local operator grows only quadratically with the number of Trotter steps, $\mathcal{O}(s^2)$. This analytical result demonstrated a specific example of the advantage of Pauli propagation in simulating local observables in models with lower operator complexity, where the computational complexity remains manageable even at late times.

Numerical benchmarks on the one-dimensional Heisenberg model confirm the theoretical results. In the free regimes ($J_z = 0$), the method attains high accuracy with a small Top-$K$ threshold $K = 2^{12}$ and significantly outperforms TDVP, which fails at longer times due to entanglement growth.
In interacting regimes ($J_z \neq 0$), the method remains competitive with TDVP, though the required K increases, reflecting the higher OSE values observed in Fig.~\ref{fig:entropy}.
The growth in the number of Pauli terms Fig.~\ref{fig:pauli_growth} and the evolution of the coefficient distribution Fig.~\ref{fig:dist} further corroborate the connection between computational cost and operator complexity as measured by the OSE.

While Pauli propagation offers a powerful alternative for simulating local observables, its performance is not universally efficient. The computational cost increases with the OSE, which is typically higher for strongly interacting systems or for non-local observables.
Future work will explore the integration of higher-order Trotter formulas to better separate truncation errors from Trotter errors, and the combination of this framework with techniques like operator-space time-evolving decimation and matrix product operator~(MPO) compression. Extensions to open quantum systems, higher dimensions, and the development of stochastic variants based on importance sampling of Pauli paths are promising directions.

In summary, the Pauli propagation method establishes an observable-centric alternative to state-based simulators for quantum dynamics simulation. It is particularly advantageous for problems where the relevant observables have low operator complexity, such as in transport studies or the calculation of out-of-time-order correlators, offering a scalable pathway in regimes where the growth of state entanglement limits traditional methods.

\section*{Author Contributions}

Y.S. developed the theoretical framework and carried out the analytical derivations.
S.C. performed the numerical simulations and analyzed the numerical results.
Z.L. and S.C. supervised the project and provided guidance.
All authors discussed the results and contributed to the writing and revision of the manuscript.

\section*{Data Availability}

All data generated and analyzed during this study, including the numerical data underlying the figures, are available from the corresponding authors upon reasonable request. This study did not use externally deposited datasets.

\begin{acknowledgments}
\emph{Acknowledgments}.---
We thank Fuchuan Wei for valuable discussions.
S.C. was supported by the National Science Foundation of China (Grant No. 12004205 and Grant No. 12574253). 
Z.L. was supported by NKPs (Grant No. 2020YFA0713000).
Y.S. and Z.L. were supported by BMSTC and ACZSP (Grant No. Z221100002722017). 
S.C. and Z.L. were supported by Beijing Natural Science Foundation (Grant No. Z220002).

\end{acknowledgments}

\bibliography{main}

\onecolumngrid
\newpage

\appendix

\section{Computational Cost}\label{ap:computational_cost}
We analyze the computational complexity of Algorithm~\ref{alg:algorithm} in terms of the number of qubits $n$, number of Trotter steps $N$, number of Pauli terms in the Hamiltonian $N_H$, and the Top-$K$ budget $K$. 

The computational cost comprises three main components:
\paragraph{Pauli conjugation}
For a term $P_i$ in Hamiltonian and $n$-qubit Pauli $P_{\alpha}$ in evolved observable, Eq.~\eqref{eq:pauli_evolution} yields:
\begin{equation}
  \begin{aligned}
    &\Big(e^{-iw_i P_i\,\tau}\Big)^\dagger P_\alpha \Big(e^{-iw_i P_i\,\tau}\Big)\\
=&
\begin{cases}
P_\alpha, & [P_i,P_\alpha]=0,\\
\cos(2w_i\tau)\,P_\alpha - i\sin(2w_i\tau)\,P_iP_\alpha, & \{P_i,P_\alpha\}=0.
\end{cases}
  \end{aligned}
\end{equation}
To implement this, we represent each Pauli operator $P_{\alpha}$ by a pair of binary strings $(\mathbf{x},\mathbf{z})\in\{0,1\}^{2n}$, where the $j$-th bits of $\mathbf{x}$ and $\mathbf{z}$ indicate whether $X$ or $Z$ acts on qubit $j$, respectively, with $Y$ represented by both bits being $1$.
The (anti)commutation relation between $P_i$ and $P_\alpha$ can be determined by calculating the symplectic inner product of their binary representations with $\order{n}$ time complexity.
If they commute, the conjugation leaves $P_\alpha$ unchanged.
If they anticommute, we compute the product $P_i P_\alpha$ by performing bitwise XOR operations on their binary strings, which also takes $\order{n}$ time.
Then, we update the coefficients according to the formula above, which takes $\order{1}$ time.
Therefore a single conjugation costs $T_{\mathrm{conj}}(n)=\order{n}$.

\paragraph{Per Hamiltonian term within one time step.}
For any time slice $s$ and a Hamiltonian term $P_i$, we need to compute the conjugation by $\Big( e^{-iw_i P_i\,\tau}\Big)^\dagger$ applied to all terms in the truncated observable $\widehat{O}_{s-1}$, followed by coefficient aggregation and Top-$K$ selection.
Because the truncated observable $\widehat{O}_{s-1}$ contains at most $K$ operators, conjugating each of these $K$ terms produces at most two output terms per input (one if commuting; a two-term linear combination if anticommuting), resulting in at most $2K$ candidate terms.
We aggregate coefficients of identical Pauli words using a hash table keyed by $(\mathbf{x},\mathbf{z})$.
Finally, we select the Top-$K$ terms by magnitude from the at most $2K$ candidates.
The costs of these three components are:
\begin{itemize}
  \item \textbf{Conjugations:} There are at most $K$ calls to the Pauli conjugation procedure, costing $\order{K\,T_{\mathrm{conj}}(n)}=\order{Kn}$.
  \item \textbf{Coefficient aggregation:} There are at most $2K$ hash insert/update operations, the cost is expected to be $\order{K}$.
  \item \textbf{Top-$K$ selection (Approximate Top-$K$):} In our implementation, we don't need to select the exact Top-$K$ terms; an approximate Top-$K$ suffices. 
  To achieve this, we use a bucket-based selection method: First, we preset $B$ magnitude buckets covering the range of possible coefficient magnitudes. 
  For example, if the maximum coefficient magnitude is $1$, we can set bucket as $[1/2, 1), [1/4, 1/2), [1/8, 1/4), \ldots, [1/2^B, 1/2^{B-1})$, for a Pauli coefficient $c$, it falls into bucket $b$ if $2^{-(b+1)} \le |c| < 2^{-b}$.
  For each candidate coefficient, we place it into the corresponding bucket in $\order{1}$ time, giving $\order{K}$ time for all $\le 2K$ candidates.
  The next step is to scan the buckets from high to low magnitude until collecting at least $K$ items, and return them as the approximate Top-$K$ set, which costs $\order{1}$ time with fixed $B$.
  Overall, the approximate Top-$K$ selection costs $\order{K}$ time.
\end{itemize}
Hence, for one $P_i$ in a time step, the total cost is $T_{\text{one }P_i}(n,K)=\order{Kn}$.

\paragraph{Per time step and total evolution.}
Each time step processes all $N_H$ Hamiltonian terms in sequence, so the cost per time step is:
\begin{equation}
  T_{\mathrm{step}}(n,K,N_H)=\order{N_H\,K\,n}.
\end{equation}
Over $N$ steps, the overall cost of the back-propagation evolution is:
\begin{equation}
  T_{\mathrm{backprop}}=\order{N\,N_H\,K\,n}.
\end{equation}

\paragraph{Final rescaling and expectation evaluation.}
After completing the back-propagation, we need to rescale the truncated observable $\widehat{O}_0$ to match the Hilbert-Schmidt norm of the original observable $\widehat{O}_0 \rightarrow \widehat{O}_0 \cdot \frac{\norm{O}_2}{\norm{\widehat{O}_0}_2}$, and then evaluate the expectation value $\langle \widehat{O} \rangle_{\rho(t)} = \tr{\widehat{O}_0 \rho(0)}$.
To compute $\frac{\norm{O}_2}{\norm{\widehat{O}_0}_2}$, assuming the norm of $O$ is pre-computed, we use the fact that the Pauli operators form an orthonormal basis under the Hilbert-Schmidt inner product:
\begin{equation}
  \norm{\widehat{O}_0}_2^2
  = \tr{\widehat{O}_0^\dagger \widehat{O}_0}
  = \tr{\left(\sum_{\alpha=1}^{K} c_\alpha P_\alpha\right)^\dagger \left(\sum_{\beta=1}^{K} c_\beta P_\beta\right)}
  = 2^n \sum_{\alpha=1}^{K} c_\alpha^2.
\end{equation}
Thus, the cost of rescaling is $\order{K}$ to obtain $\sum_{\alpha=1}^{K} c_\alpha^2$.
Next, we evaluate the expectation value $\langle \widehat{O} \rangle_{\rho(t)}$.
For a product state $\rho(0)= \otimes_{j=1}^n \rho_j$, and an expansion $\widehat{O}_0=\sum_{\alpha=1}^{K} c_\alpha P_\alpha$ with $P_\alpha=\bigotimes_{j=1}^n P_{\alpha,j}$, linearity yields
\begin{equation}
  \begin{aligned}
    \langle \widehat{O} \rangle_{\rho(t)}
    &= \tr{\widehat{O}_0\,\rho(0)} \\
    &= \sum_{\alpha=1}^{K} c_\alpha\, \tr{P_\alpha\,\rho(0)} \\
    &= \sum_{\alpha=1}^{K} c_\alpha \prod_{j=1}^n \tr{P_{\alpha,j}\,\rho_j}.
  \end{aligned}
\end{equation}
Each term $\tr{P_{\alpha,j}\,\rho_j}$ can be computed in $\order{1}$ time, so the total cost for evaluating the expectation value is $\order{K\,n}$.

Therefore, the total computational cost of Algorithm~\ref{alg:algorithm} is:
\begin{equation}
  T_{\mathrm{total}}(n,K,N_H,N) = \order{N\,N_H\,K\,n + K\,n} = \order{N\,N_H\,K\,n}.
\end{equation}

\section{Proof of Lemma~\ref{lemma:prop_entropy}}
\begin{proof}
  The proof is straightforward by using the properties of the OSE.
  \begin{enumerate}
    \item We consider the tensor product of two observables $O=O_1 \otimes O_2$, and denote the Pauli coefficients of $O$, $O_1$ and $O_2$ as $\bm{c}$, $\bm{c}_1$ and $\bm{c}_2$ respectively.
    For $P=P_1 \otimes P_2$, we have $c_P= c_{P_1} c_{P_2}$ for $P_1 \in \{I, X, Y, Z\}^{\otimes n_1}$ and $P_2 \in \{I, X, Y, Z\}^{\otimes n_2}$.
    Therefore, we have:
    \begin{equation}
      \begin{aligned}
      \norm{\bm{c^2}}^\alpha_\alpha & = \sum_{P_1 \in \{I, X, Y, Z\}^{\otimes n_1}} \sum_{P_2 \in \{I, X, Y, Z\}^{\otimes n_2}} c_{P_1 \otimes P_2}^{2\alpha}\\
      & = \sum_{P_1 \in \{I, X, Y, Z\}^{\otimes n_1}} c_{P_1}^{2\alpha} \sum_{P_2 \in \{I, X, Y, Z\}^{\otimes n_2}} c_{P_2}^{2\alpha} \\
      & = \norm{\bm{c^2}_1}^\alpha_\alpha \norm{\bm{c^2}_2}^\alpha_\alpha.
      \end{aligned}
    \end{equation}
    Taking into the entropy definition, we have:
    \begin{equation}
      \mathcal{S}^\alpha(O) = \mathcal{S}^\alpha(O_1) + \mathcal{S}^\alpha(O_2).
    \end{equation}

    \item Consider the conjugation of $O$ by a Clifford circuit $U$, we denote the Pauli coefficients of $U O U^\dagger$ as $\bm{c}'$.
    For $Q=U^\dagger P U$ with $P \in \{I, X, Y, Z\}^{\otimes n}$, by the property of the Clifford circuit, we have $Q \in \{I, X, Y, Z\}^{\otimes n}$ and $c'_Q= c_P$.
    Therefore, we have:
    \begin{equation}
      \norm{\bm{c'}^2}_\alpha = \norm{\bm{c^2}}_\alpha.
    \end{equation}
    Taking into the entropy definition, we have:
    \begin{equation}
      \mathcal{S}^\alpha(U O U^\dagger) = \mathcal{S}^\alpha(O).
    \end{equation}

    \item Let $O=\sum_i \lambda_i O_i$ be the weighted sum of the operators $O_i$.
    The Pauli coefficients of $c_i(O)$ for operator $O$ is defined as $c_i(O) = \frac{1}{2^n} \tr{P_i O}$.
    We have:
    \begin{equation}
      c_i(O) = \sum_j \lambda_j c_i(O_j),
    \end{equation}
    where $c_i(O_j) = \frac{1}{2^n} \tr{P_i O_j}$.
    Since $0 < \alpha < 0.5$, the function $f(x)=x^{2\alpha}$ is concave, by Jensen's inequality, we have:
    \begin{equation}
      \left( \sum_j \lambda_j c_i(O_j) \right)^{2 \alpha} \geq \sum_j \lambda_j c_i(O_j)^{2\alpha} .
    \end{equation}
    Therefore, we have:
    \begin{equation}
       \sum_i \left(\sum_j \lambda_j c_i(O_j)\right)^{2\alpha}  \geq   \sum_i \sum_j \lambda_j c_i(O_j)^{2\alpha} .
    \end{equation}
    Taking into the entropy definition, we have:
    \begin{equation}
      \mathcal{S}^\alpha(O) \geq \frac{1}{1-\alpha}\ln\left( \sum_i \sum_j \lambda_j c_i(O_j)^{2\alpha} \right).
    \end{equation}
    By the concavity of the logarithm function, there is:
    \begin{equation}
      \ln\left( \sum_i \sum_j \lambda_j c_i(O_j)^{2\alpha} \right) \geq \sum_i \lambda_i \ln\left( \sum_j c_i(O_j)^{2\alpha} \right).
    \end{equation}
    Therefore, we obtain $\mathcal{S}^\alpha(\cdot)$ is a concave function.
    \begin{equation}
      \mathcal{S}^\alpha(O) \geq \sum_i \lambda_i \mathcal{S}^\alpha(O_i).
    \end{equation}

    \item Similarly, we have:
    \begin{equation}
      c_i(O) = \sum_j \lambda_j c_i(O_j),
    \end{equation}
    where $c_i(O_j) = \frac{1}{2^n} \tr{P_i O_j}$.

    On the other hand, we have:
    \begin{equation}
      \exp{\mathcal{S}^\alpha(O)} =\left(\sum_i \abs{c_i(O)}^{2\alpha}\right)^{\frac{1}{1-\alpha}}=\left(\sum_i \abs{c_i(O)}^{q}\right)^{\frac{1}{1-\alpha}} =\norm{\bm{c}(O)}_q^r,
    \end{equation}
    where $q = 2\alpha$, $r = \frac{2\alpha}{1-\alpha}$ and $\bm{c}(O) = (c_1(O), c_2(O), \ldots, c_{4^n}(O))$.
    By Minkowski inequality, we have:
    \begin{equation}
      \norm{\bm{c}(O)}_q = \norm{\sum_j \lambda_j \bm{c}(O_j)}_q \leq \sum_j \lambda_j \norm{\bm{c}(O_j)}_q.
    \end{equation}

    Therefore, we have:
    \begin{equation}
       \exp{\mathcal{S}^\alpha(O)} \leq  \left( \sum_j \lambda_j \norm{\bm{c}(O_j)}_q \right)^r.
    \end{equation}

    Since $0.5 \leq \alpha <1 $, we have $r\geq 2$ and the function $f(x)=x^r$ is convex, by Jensen's inequality, we have:
    \begin{equation}
      \exp{\mathcal{S}^\alpha(O)} \leq \left( \sum_j \lambda_j \norm{\bm{c}(O_j)}_q \right)^r \leq \sum_j \lambda_j \norm{\bm{c}(O_j)}_q^r= \sum_j \lambda_j \exp{\mathcal{S}^\alpha(O_j)}.
    \end{equation}

    Therefore, we obtain $\exp{\mathcal{S}^\alpha(\cdot)}$ is a concave function.
  \end{enumerate}
\end{proof}

\section{Proof of Lemma~\ref{lemma:error_bound}}\label{ap:proof_error_bound}
\begin{proof}
  Let $\varDelta(K) \coloneqq  \sum_{i=K+1}^{\infty} c'^2_i$.
  The approximated observable is given by:
  \begin{equation}
    \widehat{O} = \frac{\norm{O}_2}{\sqrt{2^n\sum_{i=1}^{K} c'^2_i}}\sum_{i=1}^{K} c'_i P_i.
  \end{equation}
  Note that $\norm{O}_2^2 = 2^n \sum_{i=1}^{\infty} c'^2_i$, we have:
  \begin{equation}
    \norm{\widehat{O}}_2^2 = \norm{O}_2^2.
  \end{equation}
  Consider the cosine similarity between $O$ and $\widehat{O}$:
  \begin{equation}
    F_{HS}(\widehat{O},O) = \frac{\tr{O^\dagger \widehat{O}}}{\norm{O}_2 \norm{\widehat{O}}_2}=\frac{\sqrt{2^n\sum_{i=1}^{K} c'^2_i}}{\norm{\widehat{O}}_2}.
  \end{equation}
  The Hilbert-Schmidt distance can be related to the cosine similarity as:
  \begin{equation}
    \begin{aligned}
        \norm{O-\widehat{O}}_{2}^2 &= \norm{O}_2^2 + \norm{\widehat{O}}_2^2 - 2 \norm{O}_2 \norm{\widehat{O}}_2 F_{HS}(\widehat{O},O)\\
        &= 2\norm{O}_2^2 (1-F_{HS}(\widehat{O},O)).
    \end{aligned}
  \end{equation}

  On the other hand, by $2^n \sum_{i=1}^{K} c'^2_i = \norm{O}_2^2 - 2^n \varDelta(K)$,
  we have:
  \begin{equation}
    \begin{aligned}
      1-F_{HS}(\widehat{O},O) &= \frac{\norm{O}_2 - \sqrt{\norm{O}^2_2 - 2^n \varDelta(K)}}{\norm{O}_2} \\
      & = \frac{2^n \varDelta(K)}{\norm{O}_2 \left( \norm{O}_2 + \sqrt{\norm{O}^2_2 - 2^n \varDelta(K)} \right)} \\
      &\leq \frac{2^n\varDelta(K)}{\norm{O}^2_2}.
    \end{aligned}
  \end{equation}

  Therefore, the derivation can be upper bound by:
  \begin{equation}
    \norm{O-\widehat{O}}_{Pauli,2} = \frac{1}{\sqrt{2^n}}\norm{O-\widehat{O}}_2 \leq \sqrt{2\varDelta(K)}.
  \end{equation}

  For the lower bound, by $O = \sum_{i=1}^{\infty} c'_i P_i = \frac{\sqrt{2^n\sum_{i=1}^{K} c'^2_i}}{\norm{O}_2} \widehat{O} + \sum_{i=K+1}^{\infty} c'_i P_i$, we have:
  \begin{equation}
    \norm{O-\widehat{O}}_{Pauli,2} = \frac{1}{\sqrt{2^n}}\norm{O-\widehat{O}}_2 \geq \frac{1}{\sqrt{2^n}} \norm{\sum_{i=K+1}^{\infty} c'_i P_i}_2 = \sqrt{\varDelta(K)}.
  \end{equation}
  Therefore, we complete the equation:
  \begin{equation}\label{ap:eq:error_bound}
    \sqrt{\varDelta(K)} \leq \norm{O-\widehat{O}}_{Pauli,2} \leq \sqrt{2\varDelta(K)}.
  \end{equation}
\end{proof}
\begin{remark}

  \begin{enumerate}
    \item In Ref.~\cite{dowling2025magic}, a lower bound for the approximation error is given as:
  \begin{equation}
    \norm{O - \widetilde{O}}_\infty \geq \sqrt{\varDelta(K)},
  \end{equation}
  where $\widetilde{O} = \sum_{i=1}^{K} c'_i P_i$ is the truncated observable without rescaling.
  Note that by Eq.~\eqref{ap:eq:error_bound}, there is:
  \begin{equation}
    \sqrt{\varDelta(K)} \leq \norm{O - \widehat{O}}_{Pauli,2} = \frac{1}{\sqrt{2^n}}\norm{O - \widehat{O}}_2 \leq \norm{O - \widetilde{O}}_\infty.
  \end{equation}
  Therefore, for the rescaled truncated observable $\widehat{O}$, the lower bound still holds. 
  And the derivation of the lower bound is essentially the same as in Ref.~\cite{dowling2025magic}.

  \item By the properties of the operator norm, for any operator $A$, there is:
  \begin{equation}
    \frac{1}{\sqrt{4^n}}\norm{A}_\infty \leq \norm{A}_{Pauli,2} \leq \norm{A}_\infty.
  \end{equation}
  For any state $\rho$, there is:
  \begin{equation}
    \abs{\tr{A \rho}} \leq \norm{A}_\infty.
  \end{equation}
  When taking $A = O - \widehat{O}$, we can see that the $\infty$-norm gives a bound for the error of the measurement expectation value.

  While the Pauli $2$-norm gives a bound for the average error of the measurement expectation value over all pure states, i.e., using the property of Haar random states~\cite{mele2024introduction}, for traceless observable $O$, we have:
  \begin{equation}
    \mathbb{E}_{\ket{\psi} \sim \text{Haar}} \abs{\bra{\psi} \left(O - \widehat{O}\right) \ket{\psi}}^2 =  \frac{1}{4^n + 1} \norm{O - \widehat{O}}_{Pauli,2}^2,
  \end{equation}
  where the expectation is taken over the Haar measure.
  \end{enumerate}
\end{remark}

\section{Proof of Thm.~\ref{thm:sim_error_entropy}}
\begin{proof}
  First, the $\{c'^2_i\}_{i=1}^{\infty}$ is a nonincreasingly sequence, we have:
  \begin{equation}
    \abs{c'^2_i}^\alpha \leq \frac{1}{i} \sum_{j=1}^{i} \abs{c'^2_j}^\alpha \leq \frac{1}{i} \sum_{j=1}^{\infty} \abs{c'^2_j}^\alpha.
  \end{equation}
  Therefore, we have:
  \begin{equation}
    \varDelta(K) = \sum_{i=K+1}^{\infty} c'^2_i \leq \sum_{i=K+1}^{\infty} \left( \frac{1}{i} \sum_{j=1}^{\infty} \abs{c'^2_j}^\alpha \right)^{\frac{1}{\alpha}} =
    \left( \sum_{i=K+1}^{\infty} i^{-\frac{1}{\alpha}} \right) \left(\sum_{j=1}^{\infty} \abs{c'^2_j}^\alpha \right)^{\frac{1}{\alpha}}.
  \end{equation}
  On the other hand, there is :
  \begin{equation}
    \sum_{i=K+1}^{\infty} i^{-\frac{1}{\alpha}} \leq \int_{K}^{\infty} x^{-\frac{1}{\alpha}} \mathrm{d}x = \frac{\alpha}{1-\alpha} K^{\frac{\alpha-1}{\alpha}}.
  \end{equation}
  Therefore, we have:
  \begin{equation}
    \ln{\varDelta(K)} \leq \ln{\frac{\alpha}{1-\alpha}} + \frac{\alpha-1}{\alpha} \ln{K} + \frac{1}{\alpha} \ln{\left( \sum_{j=1}^{\infty} \abs{c'^2_j}^\alpha \right)} = \frac{1-\alpha}{\alpha}\left( \mathcal{S}^\alpha(O)-\ln{K} \right) + \ln{\frac{\alpha}{1-\alpha}}.
  \end{equation}
  This completes the proof of the theorem.

  By Lemma~\ref{lemma:error_bound}, we have:
  \begin{equation}
    \norm{O-\widehat{O}}_{Pauli,2} \leq \sqrt{2\varDelta(K)} \leq \sqrt{2 \frac{\alpha}{1-\alpha}} \exp{\frac{1-\alpha}{2\alpha} \left( \mathcal{S}^\alpha(O) - \ln{K} \right) }.
  \end{equation}
  To guarantee that $\norm{O-\widehat{O}}_{Pauli,2} \leq \varepsilon$, it is sufficient to choose $K$ such that:
  \begin{equation}
    K \geq \exp{\mathcal{S}^\alpha(O) } \left(\frac{2\alpha}{(1-\alpha)\varepsilon^2}\right)^{\frac{\alpha}{1-\alpha}}.
  \end{equation}
\end{proof}

\section{Proof of Thm.~\ref{thm:lower_bound}}

It's worth noting that in Ref.~\cite{dowling2025magic}, a inverse relationship between the OSE and $\Delta(K)$ was established for the case of $\alpha=1$.
\begin{equation}
  \sqrt{\Delta(K)} \geq \frac{1}{2n} \left( \mathcal{S}^{1}(O) - \log{K} - 1 \right).
\end{equation}
Here we consider the case for $\alpha \neq 1$.

\begin{proof}
  First, we consider the case that $0 < \alpha < 1$.
  Because the system is $n$ qubits, there are at most $4^n$ non-zero Pauli coefficients.
  Hence the OSE is given by:
  \begin{equation}
    \mathcal{S}^\alpha(O) = \frac{1}{1-\alpha} \ln{\left( \sum_{i=1}^{4^n} \abs{c'^2_i}^\alpha \right)},
  \end{equation}
  which is monotonically increasing with respect to $\sum_{i=1}^{4^n} \abs{c'^2_i}^\alpha$.
  For the case that $0 < \alpha < 1$, the function $f(x)=x^{\alpha}$ is concave.
  By Jensen's inequality, we have:
  \begin{equation}
    \begin{aligned}
    \sum_{i=1}^{K} \abs{c'^2_i}^\alpha &\leq \left( \frac{1}{K} \sum_{i=1}^{K} \abs{c'^2_i} \right)^{\alpha} K = \left( \norm{O}_{Pauli,2}^2 - \varDelta(K) \right)^{\alpha} K^{1-\alpha}, \\
    \sum_{i=K+1}^{4^n} \abs{c'^2_i}^\alpha &\leq \left( \frac{1}{4^n - K} \sum_{i=K+1}^{4^n} \abs{c'^2_i} \right)^{\alpha} (4^n - K) = \varDelta(K)^{\alpha} (4^n - K)^{1-\alpha},
    \end{aligned}
  \end{equation}
  where $\Delta(K) = \sum_{i=K+1}^{4^n} c'^2_i$.
  Therefore, we have:
  \begin{equation}\label{ap:eq:intermediate}
    \mathcal{S}^\alpha(O) \leq \frac{1}{1-\alpha} \ln{\left( \left( \norm{O}_{Pauli,2}^2 - \varDelta(K) \right)^{\alpha} K^{1-\alpha} + \varDelta(K)^{\alpha} (4^n - K)^{1-\alpha} \right)}.
  \end{equation}
  
  Let $g(x) = \frac{1}{1-\alpha} \ln{\left( \left( \norm{O}_{Pauli,2}^2 - x \right)^{\alpha} K^{1-\alpha} + x^{\alpha} (4^n - K)^{1-\alpha} \right)}$, we have:
  \begin{equation}
    g'(x) = \frac{\alpha}{1-\alpha} \frac{ (4^n - K)^{1-\alpha} x^{\alpha-1} - K^{1-\alpha} \left( \norm{O}_{Pauli,2}^2 - x \right)^{\alpha-1} }{ \left( \norm{O}_{Pauli,2}^2 - x \right)^{\alpha} K^{1-\alpha} + x^{\alpha} (4^n - K)^{1-\alpha} }.
  \end{equation}
  Considering the critical point $g'(x_\star)=0$, we have:
  \begin{equation}
    (4^n - K)^{1-\alpha} x_\star^{\alpha-1} - K^{1-\alpha} \left( \norm{O}_{Pauli,2}^2 - x_\star \right)^{\alpha-1} = 0,
  \end{equation}
  which gives:
  \begin{equation}
    x_\star = \norm{O}_{Pauli,2}^2 \left(1 - \frac{K}{4^n} \right).
  \end{equation}
  It's easy to verify that $g'(x)\geq 0$ when $x \leq x_\star$.

  On the other hand, we have $\sum_{i=1}^{K} c'^2_i \geq  \frac{K}{4^n} \norm{O}_{Pauli,2}^2$, which gives:
  \begin{equation}
    \varDelta(K) = \sum_{i=K+1}^{4^n} c'^2_i = \norm{O}_{Pauli,2}^2 - \sum_{i=1}^{K} c'^2_i \leq \norm{O}_{Pauli,2}^2 \left(1 - \frac{K}{4^n} \right) = x_\star.
  \end{equation}
  As a result, the right-hand side of Eq.~\eqref{ap:eq:intermediate} is a increasing function with respect to $\varDelta(K)$. 
  Thus, by Eq.~\eqref{ap:eq:intermediate}, we can obtain a implicit expression lower bound for $\varDelta(K)$ with respect to $\mathcal{S}^\alpha(O)$.

  Second, we consider the case that $\alpha > 1$.
  By Hölder's inequality, we have:
  \begin{equation}
    \left( \sum_{i=1}^{\infty} \abs{c'^2_i}^\alpha \right)^{\frac{1}{\alpha}} K^{1 - \frac{1}{\alpha}} = \left( \sum_{i=1}^{\infty} \abs{c'^2_i}^\alpha \right)^{\frac{1}{\alpha}} \left( \sum_{i=1}^{K} 1^{\frac{\alpha}{\alpha-1}} \right)^{\frac{\alpha-1}{\alpha}} \geq \sum_{i=1}^{K} \abs{c'^2_i}.
  \end{equation}
  Therefore, we have:
  \begin{equation}
    \varDelta(K) = \sum_{i=K+1}^{\infty} c'^2_i = \norm{O}_{Pauli,2}^2 - \sum_{i=1}^{K} c'^2_i \geq \norm{O}_{Pauli,2}^2 - \left( \sum_{i=1}^{\infty} \abs{c'^2_i}^\alpha \right)^{\frac{1}{\alpha}} K^{1 - \frac{1}{\alpha}}.
  \end{equation}
  Taking into the entropy definition, we have:
  \begin{equation}
    \varDelta(K) \geq \norm{O}_{Pauli,2}^2 - \exp{\frac{1-\alpha}{\alpha} \mathcal{S}^\alpha(O)} K^{1 - \frac{1}{\alpha}}.
  \end{equation}
  Therefore, we can obtain:
  \begin{equation}
    \ln{K} - \frac{\alpha}{\alpha-1} \ln{\left( \norm{O}_{Pauli,2}^2 - \varDelta(K) \right)} \geq \mathcal{S}^\alpha(O).
  \end{equation}
\end{proof}

\begin{remark}
    Now, we can take $\alpha \to 1_-$ to recover the result in Ref.~\cite{dowling2025magic}, before taking the limit, we add the limitation in Ref.~\cite{dowling2025magic} that $\norm{O}_{Pauli,2}^2 = 1$, using Eq.~\eqref{ap:eq:intermediate}, we have:
  \begin{equation}
    \begin{aligned}
      \mathcal{S}^{1}(O) = \lim_{\alpha \to 1_-} \mathcal{S}^\alpha(O)  & \leq \lim_{\alpha \to 1_-} \frac{1}{1-\alpha} \ln{\left( \left( 1 - \varDelta(K) \right)^{\alpha} K^{1-\alpha} + \varDelta(K)^{\alpha} (4^n - K)^{1-\alpha} \right)} \\
      & = \lim_{\alpha \to 1_-} - \frac{\left( \left( 1 - \varDelta(K) \right)^{\alpha} K^{1-\alpha} + \varDelta(K)^{\alpha} (4^n - K)^{1-\alpha} \right)'}{\left( \left( 1 - \varDelta(K) \right)^{\alpha} K^{1-\alpha} + \varDelta(K)^{\alpha} (4^n - K)^{1-\alpha} \right)} \\
      & = - (1 - \varDelta(K)) \ln{\frac{1 - \varDelta(K)}{K}} - \varDelta(K) \ln{\frac{4^n - K}{\varDelta(K)}}\\
      & = -(1 - \varDelta(K)) \ln{(1 - \varDelta(K))} - \varDelta(K) \ln{\varDelta(K)} + (1 - \varDelta(K)) \ln{K} + \varDelta(K) \ln{(4^n - K)} \\
      & \leq \ln{2} + \ln{K} + \varDelta(K) 2n \ln{2},
    \end{aligned}
  \end{equation}
  where the last inequality is due to the fact that the binary entropy function $-x \ln{x} - (1-x) \ln{(1-x)} \leq \ln{2}$ for $0 \leq x \leq 1$ and $0 \leq \varDelta(K) \leq 1$.
  Rearranging the terms, we have:
  \begin{equation}
    \sqrt{\varDelta(K)} \geq \varDelta(K) \geq \frac{1}{2n \ln{2}} \left( \mathcal{S}^{1}(O) - \ln{K} - \ln{2} \right) = \frac{1}{2n} \left( \frac{1}{\ln{2}}\mathcal{S}^{1}(O) - \log{K} - 1 \right).
  \end{equation}
  This recovers the result in Ref.~\cite{dowling2025magic} up to a constant factor $\frac{1}{\ln{2}}$ due to the different logarithm bases used in the definition of the OSE.
\end{remark}

\section{Proof of Thm.~\ref{thm:1d_XY_trotter}}
\begin{proof}
  The Trotterized evolution employs the following gates:
  \begin{equation}
    U_{j,j+1}(\theta_1) = \exp{i\theta X_j X_{j+1}},
  \end{equation}
  and
  \begin{equation}
    V_{j,j+1}(\theta_2) = \exp{i\theta Y_j Y_{j+1}},
  \end{equation}
  where $\theta_1 = -J_X \tau$ and $\theta_2 = -J_Y \tau$, respectively.
  
  Define the operators $J_{i,j}=X_i Z_{i+1} \ldots Z_{j-1}Y_j$, $H_{i,j}=Y_i Z_{i+1} \ldots Z_{j-1}X_j$, $K_{i,j}=X_i Z_{i+1} \ldots Z_{j-1}X_j$ and $L_{i,j}=Y_i Z_{i+1} \ldots Z_{j-1}Y_j$. 
  The subspace:
  \begin{equation}
    {span} \left\{Z_l, J_{l,m}, H_{l,m}, K_{l,m}, L_{l,m} \right\}
  \end{equation}
  is invariant under the conjugation by $U_{j,j+1}(\theta)$ and $V_{j,j+1}(\theta)$.

  For example, conjugating $Z_l$ by $U_{j,j+1}(\theta_1)$, we have:
  \begin{equation}
    \begin{aligned}
      &U_{j,j+1}(\theta_1)^\dagger Z_l U_{j,j+1}(\theta_1) \\
      &= 
    \begin{cases}
      Z_l, & l \neq j, j+1 \\
      \cos(2\theta_1) Z_j - \sin(2\theta_1) H_{j,j+1}, & l=j \\
      \cos(2\theta_1) Z_{j+1} - \sin(2\theta_1) J_{j,j+1}, & l=j+1
    \end{cases}
    \end{aligned}
  \end{equation}
  which confirms the claimed invariance.

  For other operators in the algebra, the conjugation action can be verified similarly.

  On the other hand, after $s$-step Trotterization, the light-cone of $Z_l$ is contained in the interval $[l-s, l+s]$.
  Consequently, the number of non-zero Pauli coefficients in the evolved operator $Z_l$ is upper bounded by:
  \begin{equation}
    \binom{2s}{1} + 4 \binom{2s}{2} = 2s + 4s(2s-1) = \order{s^2},
  \end{equation}
  where $\binom{2s}{1}$ counts the number of $Z_i$ terms, and $4 \binom{2s}{2}$ counts the number of $J_{i,j}, H_{i,j}, K_{i,j}, L_{i,j}$ terms.

  Therefore, the OSE of the evolved operator $Z_l$ is upper bounded by:
  \begin{equation}
    \mathcal{S}^\alpha(U_s^\dagger Z_l U_s) = \frac{1}{1-\alpha} \ln{\left( \sum_{i=1}^{\infty} \abs{c'^2_i}^\alpha \right)} \leq \frac{1}{1-\alpha} \ln{N \cdot \frac{1}{N^\alpha}}  = \ln{N} = \order{\ln{s}},
  \end{equation}
  where $N = \order{s^2}$ is the number of non-zero Pauli coefficients in the evolved operator $Z_l$ and we have used the fact that the OSE takes the maximum value $\ln{N}$ when all the non-zero Pauli coefficients have the same magnitude, i.e., $c'^2_i = \frac{1}{N}$ for $i=1,2,\ldots,N$.
  
\end{proof}

\section{Weight Truncation}\label{ap:weight_truncation}
For weight truncation, previous works~\cite{angrisani2024classically} have established error bounds under structural assumptions such as local scrambling.
We assume the state $\rho$ is sampled from a local-scrambling ensemble, i.e $\rho = V\ketbra{0}{0}^{\otimes n}V^\dagger$ with $V$ sampled from a local-scrambling ensemble.
Then, the expectation error is bounded by:
\begin{theorem}\label{thm:local_scrambling}
  Let $O = \sum_{P \in \{I, X, Y, Z\}^{\otimes n}} c_P P$ be an observable and $\widehat{O}$ be the approximated observable obtained by weight truncation with threshold $M$ and norm rescaling:
  \begin{equation}
    \widehat{O} = \frac{\norm{O}_2}{\sqrt{2^n\sum_{wt(P) < M} c_P^2}}\sum_{wt(P) < M} c_P P,
  \end{equation}
  where $wt(P)$ is the Hamming weight of the Pauli word $P$, i.e., the number of non-identity Paulis in $P$.
  Then for a state $\rho$ sampled from a local-scrambling ensemble, we have:
  \begin{equation}
    \mathbb{E}_{\rho} \abs{\langle \widehat{O} \rangle_{\rho} - \langle O \rangle_{\rho}}^2 \leq \left(\frac{2}{3}\right)^M \norm{O}_{Pauli,2}^2 + \sum_{wt(P) \geq M} c_P^2,
  \end{equation}
  where $\norm{O}_{Pauli,2} = \sqrt{\sum_P c^2_P}$ is normalized Hilbert-Schmidt norm.
\end{theorem}
This result is similar to that in Ref.~\cite{angrisani2024classically}, except that there is a rescaling procedure after truncation.

In our numerical experiments, we observe that, in 1D heisenberg chain, the weight truncation strategy is nonessential, since under the Top-$K$ truncation only lower weights usually survive.

\begin{figure}[htbp]
  \centering
  \includegraphics[width=0.7\textwidth]{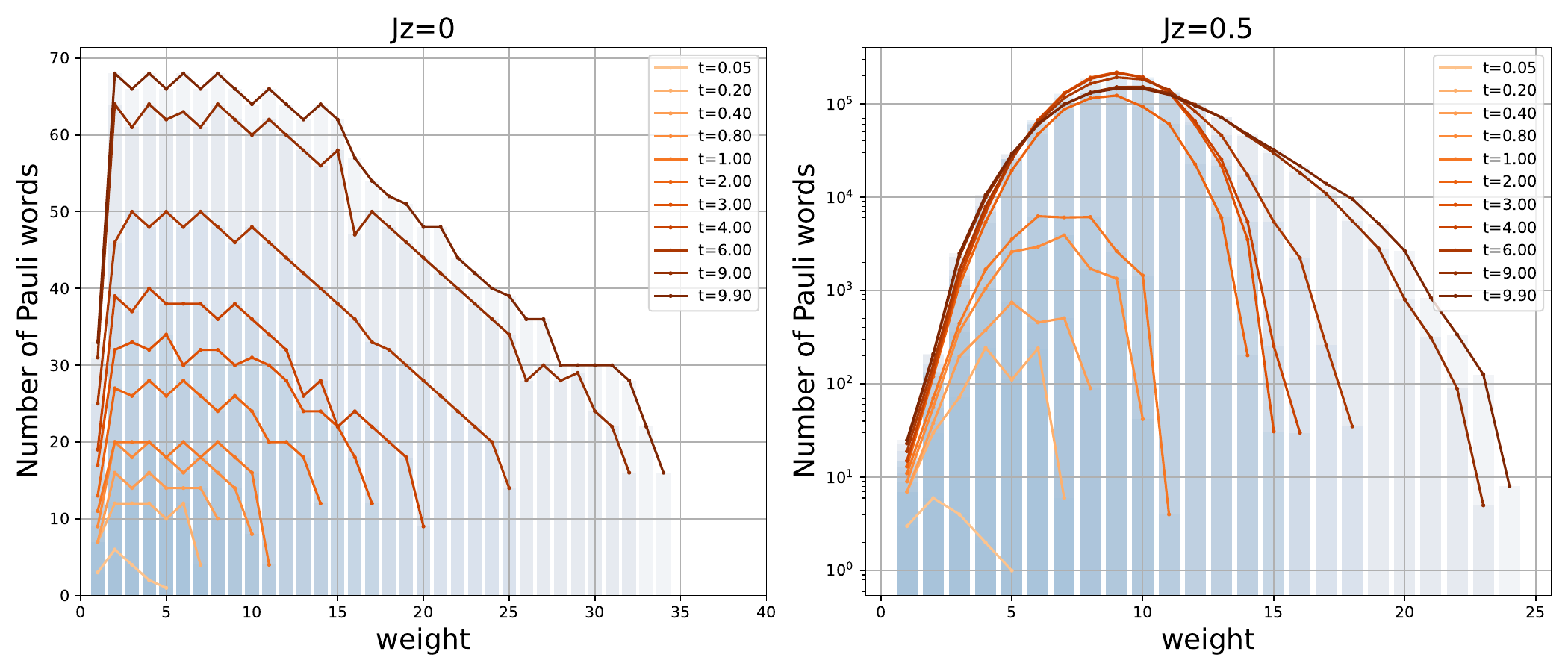}
  \caption{Pauli weight distribution of the Pauli words coefficients as a function of time for different values of $J_z$.
  }
  \label{fig:weight}
\end{figure}

As shown in Fig.~\ref{fig:weight}, we analyze the distribution of the Pauli words coefficients in terms of their Hamming weight.
We can see that the distribution of the Pauli words coefficients becomes more dispersed with time, and the dispersion becomes more significant with the increase of $J_z$. This indicates that the strategy of deactivate weight truncation is effective.
Moreover, we could observe that as time evolves, the operator becomes more complex and involves more qubits, and the increase of $J_z$ further enhances this effect.

\subsection{Proof of Theorem~\ref{thm:local_scrambling}}
\begin{proof}
  The proof is similar to that in Ref.~\cite{angrisani2024classically}, except that there is a rescaling procedure after truncation.
  By $\rho=V\ketbra{0}{0}^{\otimes n}V^\dagger$ with $V$ sampled from a local-scrambling ensemble, we have:
  \begin{equation}
    \mathbb{E}_{\rho} \abs{\langle \widehat{O} \rangle_{\rho} - \langle O \rangle_{\rho}}^2 = \mathbb{E}_{V} \abs{\tr{V^\dagger (\widehat{O}-O) V \ketbra{0}{0}^{\otimes n}}}^2. 
  \end{equation}

  The difference between the approximated observable and the original observable is given by:
  \begin{equation}
    \begin{aligned}
      &\widehat{O}-O=  \\
      &\left(\frac{\norm{O}_2}{\sqrt{2^n\sum_{wt(P) < M} c_P^2}} - 1\right) \left( \sum_{wt(P) < M} c_P P \right) \\
      &- \left( \sum_{wt(P) \geq M} c_P P \right).
    \end{aligned}
  \end{equation}
  
  Using the property of local-scrambling ensemble (Lemma~7 in Ref.~\cite{angrisani2024classically}), we have:
  \begin{equation}
    \begin{aligned}
      &\mathbb{E}_{V} \tr{V^\dagger P V \ketbra{0}{0}^{\otimes n}}\tr{V^\dagger Q V \ketbra{0}{0}^{\otimes n}} \\
      &= \begin{cases}
        0, & P \neq Q, \\
        \frac{1}{3^{wt(P)}} \sum_{ W} 
        \mathbb{E}_{V} \tr{V^\dagger W V \ketbra{0}{0}^{\otimes n}}^2
        , & P = Q,
      \end{cases}
    \end{aligned}
  \end{equation}
  where the summation of $W$ is over all Pauli words with same support as $P$, i.e., $\supp (W) = \supp (P)$ for $
        W\in \{I, X, Y, Z\}^{\otimes n}$.

  On the other hand, by Lemma 10 in Ref.~\cite{angrisani2024classically}, there is:
  \begin{equation}
    \mathbb{E}_{V} \tr{V^\dagger W V \ketbra{0}{0}^{\otimes n}}^2 \leq \left( \frac{2}{3} \right)^{wt(W)}.
  \end{equation}

  Combining the above equations, we have:
  \begin{equation}
    \begin{aligned}
      &\mathbb{E}_{V} \abs{\tr{V^\dagger (\widehat{O}-O) V \ketbra{0}{0}^{\otimes n}}}^2 \\
      &\leq\left(\frac{\norm{O}_2}{\sqrt{2^n\sum_{wt(P) < M} c_P^2}} - 1\right)^2 \left( \sum_{wt(P) < M}  c_P^2\right) \\
      &+ \left( \sum_{wt(P) \geq M} \left( \frac{2}{3} \right)^{wt(P)} c_P^2 \right)
    \end{aligned}
  \end{equation}

  Using $\norm{O}_2^2 = 2^n \sum_{P} c_P^2$, we have:
  \begin{equation}
    \begin{aligned}
      &\left(\frac{\norm{O}_2}{\sqrt{2^n\sum_{wt(P) < M} c_P^2}} - 1\right)^2 \\
      &=\left(\frac{\sqrt{2^n\sum_{P}c_P^2}}{\sqrt{2^n\sum_{wt(P) < M} c_P^2}} - 1\right)^2 \\
    & \leq \frac{\sum_{wt(P) \geq M} c_P^2}{\sum_{wt(P) < M} c_P^2}.
    \end{aligned}
  \end{equation}

  Therefore, we have:
  \begin{equation}
    \mathbb{E}_{\rho} \abs{\langle \widehat{O} \rangle_{\rho} - \langle O \rangle_{\rho}}^2 \leq \left(\frac{2}{3}\right)^M \norm{O}_{Pauli,2}^2 + \sum_{wt(P) \geq M} c_P^2.
  \end{equation}
  \end{proof}

\section{Computational complexity versus Jz perturbation}

\begin{figure}[htbp]
  \centering
  \includegraphics[width=0.6\textwidth]{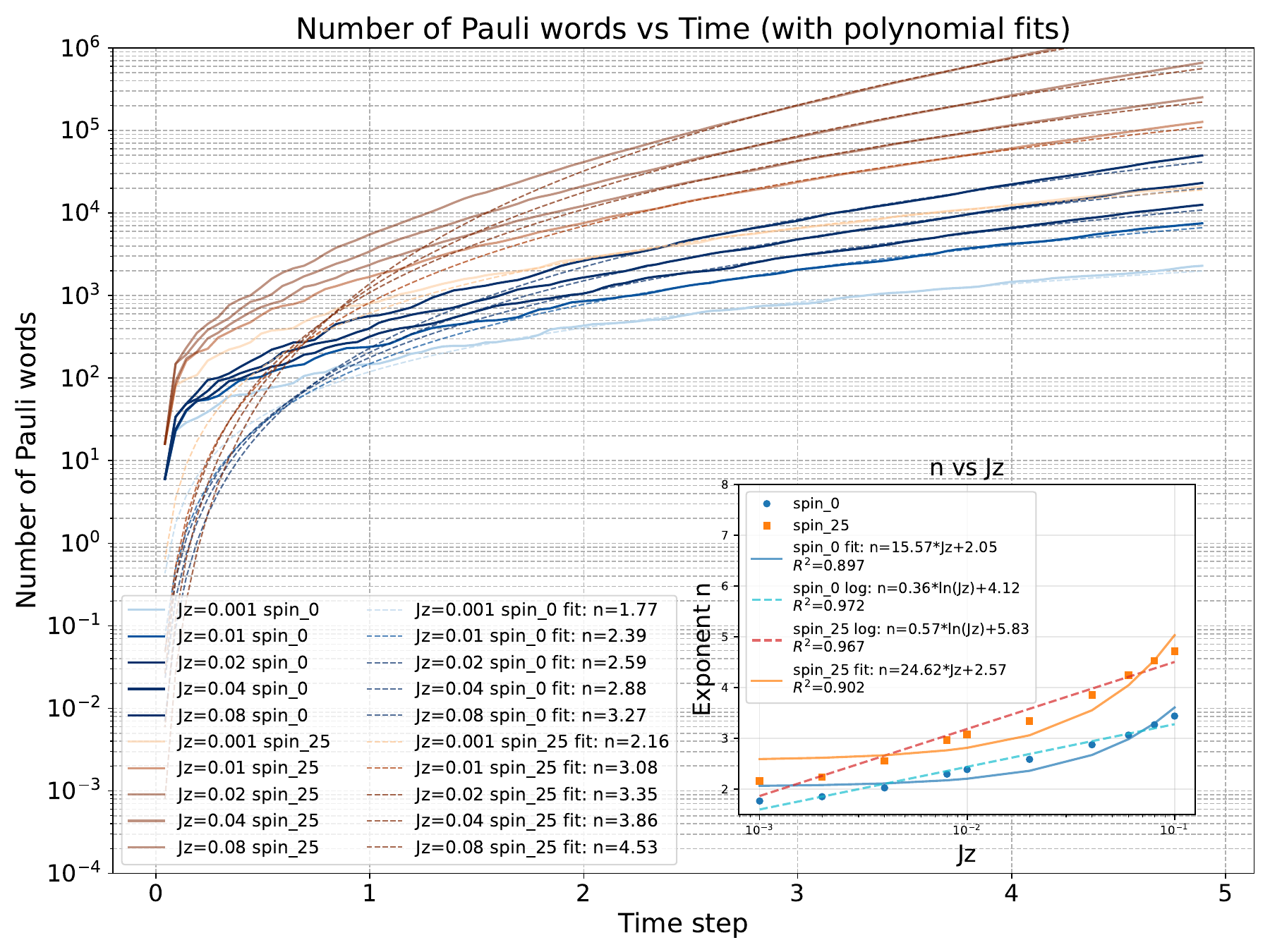}
  \caption{Growth of the number of non-zero Pauli coefficients over time for small $J_z$. The strictly linear growth at $J_z=0$ is lifted to a polynomial growth for non-zero perturbations.}
  \label{fig:layersize_compare}
\end{figure}

In this section, we investigate the impact of $J_z$ perturbations on the computational complexity of Pauli propagation method for 1D XXZ model. We first evaluate the impact of small $J_z$ on the number of non-zero Pauli coefficients with magnitude greater than $10^{-8}$. 
Due to this threshold, the estimated count is significantly smaller than the number of non-zero elements in the Dynamical Lie Algebra (DLA), but it provides a more realistic estimate of the actual algorithmic computational complexity. 

As shown in Fig.~\ref{fig:layersize_compare}, we plot the results for different $J_z$ values. 
We performed an exponential fit $f(t)=c e^{ \eta t}$ on the curves using data from $t \in [1,4]$. 
The inset illustrates the relationship between the growth exponent $\eta$ and $J_z$, revealing a clear linear relationship between $\eta$ and $\log(J_z)$ in the vicinity of small $J_z$.
For the strictly free case $J_z=0$, the number of non-zero Pauli coefficients grows linearly, consistent with the prediction of Theorem~\ref{thm:1d_XY_trotter}. 
For non-zero $J_z$, the growth exhibits an exponential trend, which becomes more pronounced as $J_z$ increases. 
This behavior suggests that while the system remains simulatable for short to intermediate times under small perturbations, the long-term computational cost eventually diverges from the efficient $J_z=0$ limit.

\begin{figure}[htbp]
  \centering
  \includegraphics[width=0.7\textwidth]{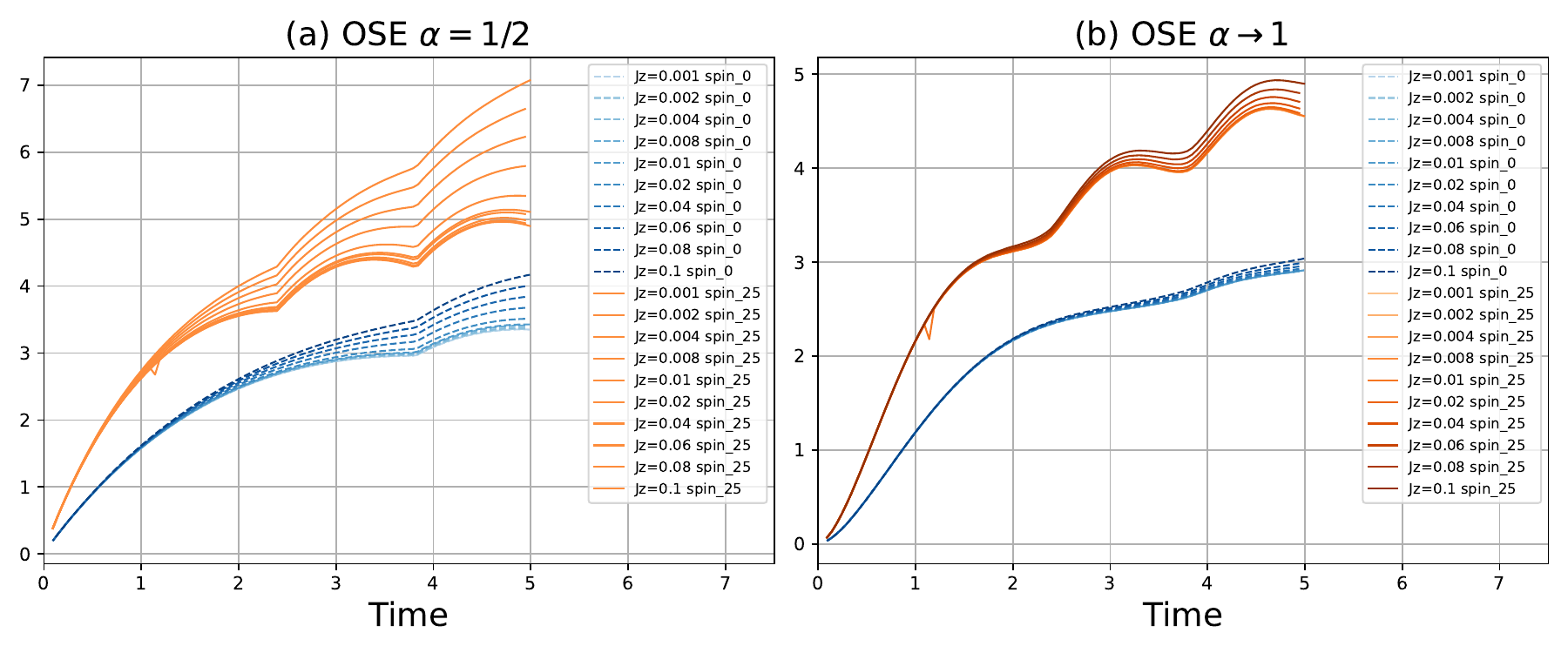}
  \caption{Time evolution of the $\alpha=1/2$ Rényi OSE for small $J_z$ perturbations. The entropy increases monotonically with both time and the coupling strength $J_z$, quantitatively capturing the rise in operator complexity as the system deviates from the free-fermion point ($J_z=0$).}
  \label{fig:entropy_compare_half}
\end{figure}

\begin{figure}[htbp]
  \centering
  \includegraphics[width=0.7\textwidth]{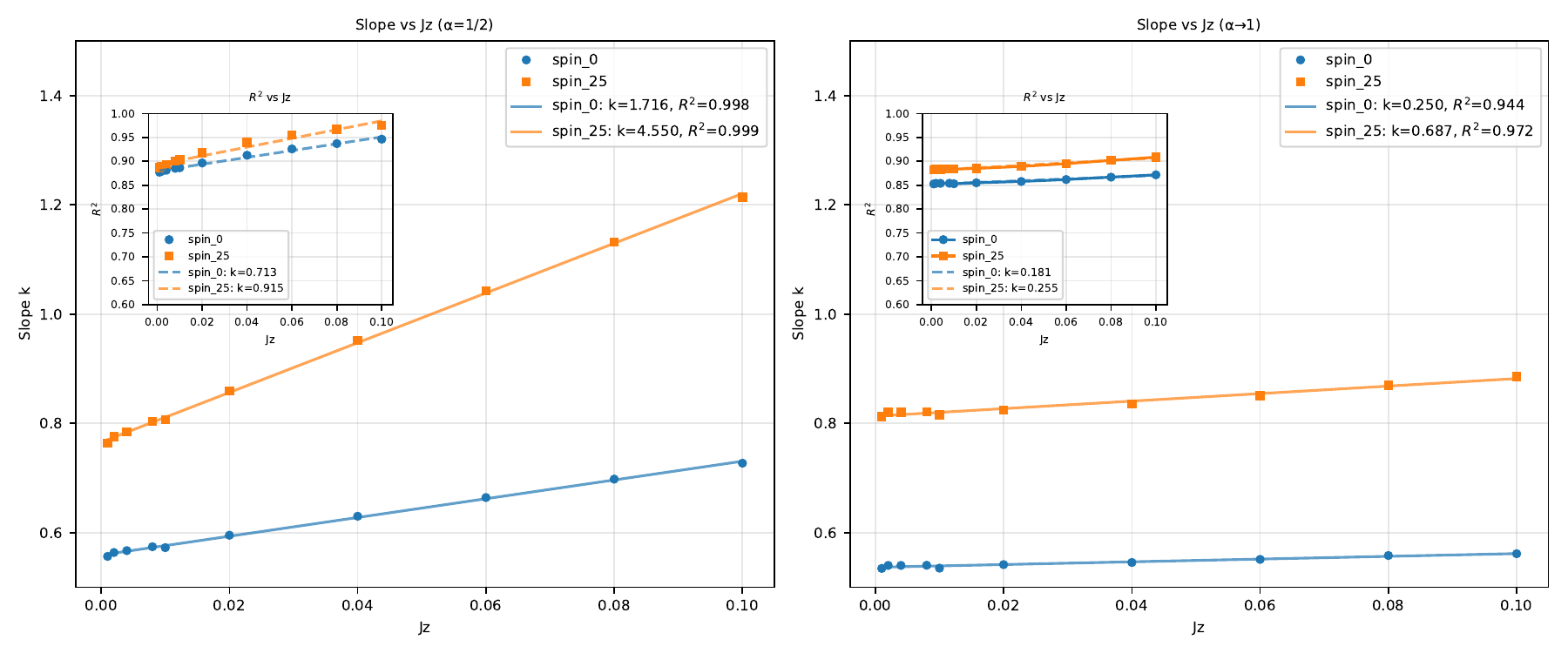}
  \caption{Time evolution of the Shannon OSE ($\alpha \to 1$) for small $J_z$ perturbations. Similar to the $\alpha=1/2$ case, the Shannon entropy acts as a sensitive probe for complexity growth, indicating that even minor interactions lead to a discernible increase in computational resource requirements.}
  \label{fig:entropy_compare_slope}
\end{figure}

Moreover, we provide a detailed visualization of the entropic growth in Fig.~\ref{fig:entropy_compare_half} and Fig.~\ref{fig:entropy_compare_slope}.
Fig.~\ref{fig:entropy_compare_half} presents ,under variant $J_z$ perturbations, the time evolution of the $\alpha=1/2$ Rényi OSE and the Shannon OSE ($\alpha \to 1$).
Consistent with our previous observations, the entropy increases monotonically with both time and the coupling strength $J_z$. Even a small perturbation from the free-fermion point ($J_z=0$) leads to a discernible increase in operator complexity, implying a higher demand for computational resources to maintain accuracy.

These plots reinforce the conclusion that operator complexity serves as a sensitive probe for interactions, with the entropy growth rate increasing clearly alongside the perturbation strength $J_z$.

In Fig.~\ref{fig:entropy_compare_slope}, we perform a linear fit $S(t)=kt+c$ on the curves from Fig.~\ref{fig:entropy_compare_half} in the interval $t \in [1,4]$, and illustrate the relationship between the linear slope $k$ and $J_z$. The results show that $k$ exhibits a linear relationship with $J_z$ in the small $J_z$ regime. This indicates that even minute interactions can lead to a significant increase in the rate of complexity growth. The inset displays the confidence of the linear fits for different $J_z$ values.

\section{Comparison to other variants of Pauli backpropagation}\label{ap:comparison}

In this section, we provide a concise comparison between variants of Pauli backpropagation methods proposed in recent works, particularly focusing on their truncation schemes, noise models, and applicable circuit types. 
The comparison is summarized in Table~\ref{tab:table_pauli}.

\begin{table*}[h!]
\centering
\begin{ruledtabular}
\begin{tabular}{|p{1.8cm}||p{2.0cm}|p{1.8cm}|p{1.8cm}|p{1.8cm}|p{1.8cm}|p{1.8cm}|p{1.8cm}|p{1.8cm}|}
%\hline
Reference & OBPPP~\cite{shao2024simulating} LOWESA~\cite{fontana2025classical} & Ref. \cite{schuster2024polynomial} & Ref. \cite{gonzalez2025pauli} & Ref. \cite{angrisani2024classically} & SPD~\cite{beguvsic2024fast,beguvsic2025simulating} (xSPD~\cite{beguvsic2025real}) & Ref. \cite{martinez2025efficient} & Ref. \cite{angrisani2025simulating} & Ref.~\cite{loizeau2025quantum,Loizeau2025} \& This work \\ \hline
Truncation scheme & Total Hamming weight & Layer-wise Hamming weight & Total Hamming weight & Layer-wise Hamming weight & Number of sinne (as well as X weight)& Tree depth & Total Hamming weight & Top-$K$ coefficient truncation with rescaling \\ \hline
Noise model & Unital noise  & Depolarizing + random amplitude damping & Depolarizing & Noiseless & Noiseless & Non-unital noise & Arbitrary local noise & Noiseless \\ \hline
Circuit & Any circuit, random angles & Any circuit, random input states& Clifford + T & Random circuits & Any circuit, small angles & (Almost) any circuit, random angles & Random circuits & Low OSE circuits \\ \hline
%\hline
\end{tabular}
\end{ruledtabular}
\caption{Comparison of the different Pauli backpropagation methods and their guarantees. The runtime may depend on the number of qubits $n$, the depth of the quantum circuit $m$, the circuit geometry $D$, the additive precision $\epsilon$ and the noise strength $p$.}
\label{tab:table_pauli}
\end{table*}

Additionally, there are some other works that focus on the Software implementation ~\cite{rudolph2025pauli,loizeau2025quantum}.

%\nocite{*}
%\bibliographystyle{apsrev}

\end{document}